\documentclass[aps,pre,superscriptaddress,onecolumn]{revtex4-2}

\usepackage{graphicx,amsmath,amssymb}

\newcommand{\dd}{\text{d}}
\newcommand{\He}{\text{He}}
\newcommand{\erf}{\text{erf}}
\newcommand{\erfc}{\text{erfc}}

\begin{document}

\title{Discrete sampling of correlated random variables \\ modifies the long-time behavior of their extreme value statistics}

\author{Lior Zarfaty}
\affiliation{Department of Physics, Institute of Nanotechnology and Advanced Materials, Bar-Ilan University, Ramat-Gan 52900, Israel}

\author{Eli Barkai}
\affiliation{Department of Physics, Institute of Nanotechnology and Advanced Materials, Bar-Ilan University, Ramat-Gan 52900, Israel}

\author{David A. Kessler}
\affiliation{Department of Physics, Bar-Ilan University, Ramat-Gan 52900, Israel}

\begin{abstract}

We consider the extreme value statistics of correlated random variables that arise from a Langevin equation. Recently, it was shown that the extreme values of the Ornstein-Uhlenbeck process follow a different distribution than those originating from its equilibrium measure, composed of independent and identically distributed Gaussian random variables. Here, we first focus on the discretely sampled Ornstein-Uhlenbeck process, which interpolates between these two limits. We show that in the limit of large times, its extreme values converge to those of the equilibrium distribution, instead of those of the continuously sampled process. This finding folds for any positive sampling interval, with an abrupt transition at zero. We then analyze the Langevin equation for any force that gives rise to a stable equilibrium distribution. For forces which asymptotically grow with the distance from the equilibrium point, the above conclusion continues to hold, and the extreme values for large times correspond to those of independent variables drawn from the equilibrium distribution. However, for forces which asymptotically decay to zero with the distance, the discretely sampled extreme value statistics at large times approach those of the continuously sampled process.

\end{abstract}

\maketitle

\section{Introduction}

Extreme value (EV) statistics is a long existing field of probability theory \cite{Gumbel,Leadbetter}, which has drawn much interest over the years, having applications in various fields of science, see for example Refs.~\cite{Sornette,Mikosch,Krapivsky,Dean,Mori}, among others. It deals with the distribution of EVs of random variables (RV), and in a similar way to the central limit theorems for sums of RVs, there are various limit laws that apply when the sample size $N$ approaches infinity. For independent and identically distributed (IID) RVs, the EV cumulative distribution function (CDF) has an exact solution in terms of $F^{\rm I}(z)$, the CDF of the underlying distribution, namely $F^{\rm I}_N(z)=[F^{\rm I}(z)]^N$. This immediately implies the existence of a limiting form for the EV distribution when $N\to\infty$. For underlying probability density functions (PDF) that fall off faster than any power-law, e.g. the Gaussian, this limit is a Gumbel distribution, $F^{\rm I}_N(z) \simeq \exp[-\exp(-z_{\rm r})]$, in the shifted and rescaled variable $z_{\rm r} \equiv (z-b_N)/a_N$, with $a_N$ and $b_N$ scaling sequences \cite{Fisher,Gnedenko,Hall}. This IID case has been studied in numerous works in the past decades, see Refs.~\cite{Sanjib,Gyorgyi,Rita,Fortin,Zarfaty}, to name a few. However, the EV statistics of correlated RVs still remains largely unexplored \cite{Grebenkov,DeBruyne}.

In a recent work by Majumdar, et al. \cite{Pal,Majumdar}, the EV statistics of correlated RVs governed by a Langevin equation were considered \citep[see also][]{Godec,Kearney}. In particular, they showed that at long times $T$, the CDF of the EV distribution of the continuous-time Ornstein-Uhlenbeck (OU) process takes the form of a $T$-independent function $F^{\rm c}(z)$ raised to the $T$th power, $F^{\rm c}_T(z)\simeq[F^{\rm c}(z)]^{T/\tau}$, where $\tau$ is the relaxation time of the process. This is of course very reminiscent of the form of the EV CDF for IID variables, with $T/\tau$ now playing the role of $N$, the number of independent samples. Moreover, since $1-F^{\rm c}(z)$ has a Gaussian decay for large $z$, one immediately learns the important result that this correlated EV distribution also converges to the universal Gumbel form in the large-$T$ limit.

\begin{figure}
	\includegraphics[width=0.75\textwidth]{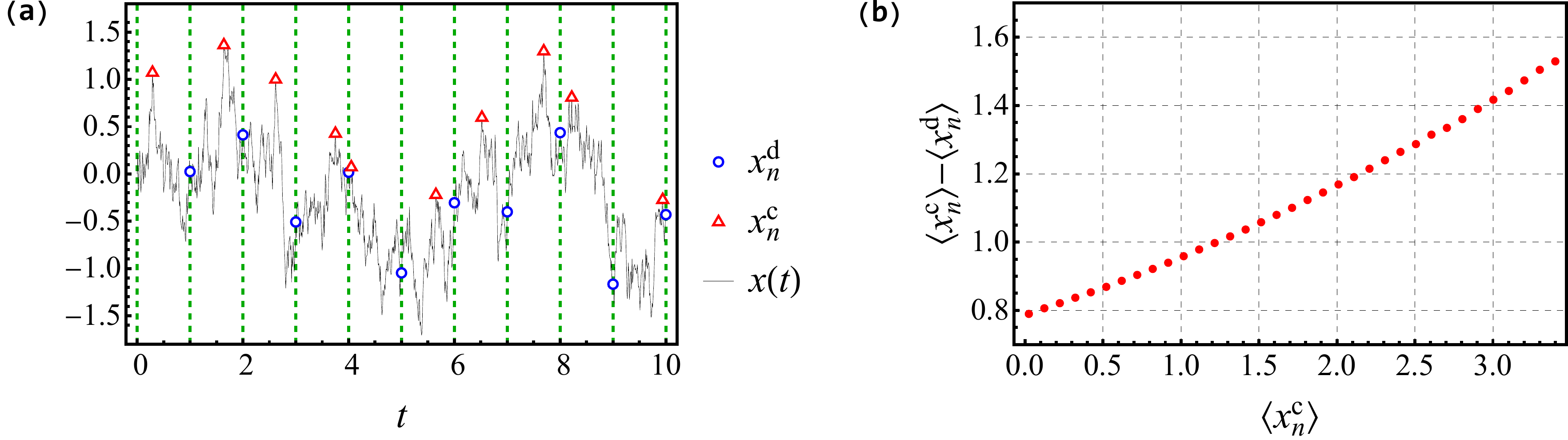}
	\caption{(Color online) {\bf The naive argument} suggests that for a large measurement time $T$, the OU time series $x(t)$ should contain approximately $T/\tau$ independent samples drawn from the equilibrium Gaussian distribution ($\tau$ is the correlation time). (a) The trajectory of the OU process, $x(t)$, ranged over the interval $[0,T=10]$, can be divided into blocks that are of order of a correlation time $\tau=1$ in size (separated by the vertical green dashed lines). Defining $x^{\rm d}_n$ as the value of $x(t)$ at the end of the $n$th interval and $x^{\rm c}_n$ as the maximum of $x(t)$ over the $n$th interval, we clearly see that $x^{\rm d}_n \le x^{\rm c}_n$ for all $1\le n\le N=T/\tau$. (b) However, the means of $x^{\rm c}_n$ and $x^{\rm d}_n$ are clearly correlated, supporting the naive argument. Nevertheless, the correlation weakens as $\langle x_n^{\rm c}\rangle$ increases, invalidating the naive argument. $10^3$ realizations of trajectories of length $T=10^5$ were generated, and the values of the blocks were binned by percentiles. The initial condition for all trajectories was $x(0)=0$.}
\label{figure: ou trajectory}
\end{figure}

Naively, one might expect that for large $T$ the OU time series $x(t)$ will contain of order $T/\tau$ independent samples, drawn from the process' equilibrium distribution, which is a Gaussian, but this turns out not to be the case. To see more clearly the logic underlying this expectation, and why it is misleading, we plot a trajectory sample of the OU process in Fig.~\ref{figure: ou trajectory}(a), with $T=10$ and $\tau=1$. Exploiting the argument presented by Majumdar, et al. \cite{Majumdar}, we split the trajectory into $N=T/\tau=10$ blocks that are roughly uncorrelated between themselves, which explains the shape of $F^{\rm c}_T(z)$ described above. Let us denote $x^{\rm d}_n \equiv x(n\tau)$, the value of $x(t)$ at the end of the $n$th interval (denoted by the blue circles in the figure) and $x^{\rm c}_n \equiv \max_{(n-1)\tau<t\le n\tau}[x(t)]$, the maximum of $x(t)$ over the $n$th interval (denoted by the red triangles), where $1\le n\le N$. It is clear that the EV $z \equiv \max_{0< t\le T}[x(t)] = \max_{1\le n\le N}(x^{\rm c}_n)$. Obviously, for all $n$ one has $x^{\rm d}_n \le x^{\rm c}_n$, as can be seen in Fig.~\ref{figure: ou trajectory}(a). Moreover, to the extent that the $x^{\rm d}_n$ are uncorrelated, the distribution of the $x^{\rm d}_n$s is just that of IID Gaussian variables. From this we deduce that the underlying CDF of the $x^{\rm c}_n$s, $F^{\rm c}$, must differ from that of the $x^{\rm d}_n$s, $F^{\rm I}$, which is a Gaussian as mentioned.

Nevertheless, for any given $n$, $x^{\rm c}_n$ and $x^{\rm d}_n$ are correlated, since they are separated by a time of only $\tau/2$ on average. We can verify this by plotting in Fig.~\ref{figure: ou trajectory}(b) $\langle x^{\rm c}_n\rangle-\langle x^{\rm d}_n\rangle$ as a function of $\langle x^{\rm c}_n\rangle$, where $\langle\cdot\rangle$ denotes mean over realizations. If the two data sets were uncorrelated, the mean of $x^{\rm d}_n$ would be zero, and we would have a straight line with slope $1$. We see that the slope of the curve is much smaller than unity, showing that the two data sets are indeed correlated. But if the two data sets track each other, and the EV statistics of $x^{\rm d}_n$ is roughly that of IID Gaussian variates, we would expect that the EV statistics of $x^{\rm c}_n$ would behave similarly. The results of Majumdar, et al. show however that they do not. In fact, as we will discuss in detail, the quantities $1-F^{\rm c}(z)$ and $1-F^{\rm I}(z)$, while sharing the same Gaussian falloff, differ by a factor that grows as $z^2$. This is related to the fact that the slope of the curve in Fig.~\ref{figure: ou trajectory}(b) increases with $\langle x^{\rm c}_n \rangle$, showing that for large values the correlation gets weaker. The difference in prefactors leads to the following anomaly. One can, for any interval $T$, define $N_{\rm eff}(T)$ as the number of Gaussian IID variables that would need to be drawn to give the same mean $z$ as in the continuously-sampled case. It turns out that at large $T$, $N_{\rm eff}(T)$ is not proportional to $T$, but rather $N_{\rm eff}(T)/T$ diverges with $T$.

In this work, we explore this puzzling difference between the continuously-sampled and the IID cases by noting that the IID case can be considered as a discrete sampling of the continuous process, with a long interval between samplings. The problem of discrete sampling naturally arises in many cases, since in any experiment one always has a minimal measurement interval. This immediately raises two questions: I) What are the statistics of discretely sampled processes? Moreover, II) What is the nature of the transition from an IID system to the continuously-sampled one, as one varies the sampling time?

We start addressing these questions by investigating the discretely sampled OU process. Given a sampling time of $\Delta\ge0$, we find that the aforementioned transition occurs sharply at $\Delta=0$. Namely, we find that for any $\Delta>0$ and large enough $T$, the EV statistics is of $T/\Delta$ IID RVs drawn from the equilibrium Gaussian PDF $\exp(-z^2/2)/\sqrt{2\pi}$. After providing a simple explanation for this phenomenon, we expand our inquiry to more general Langevin processes. We extend Majumdar's approach and calculate the EV statistics for a continuously sampled process with potentials which grow faster than linearly with distance. Our findings suggest that asymptotically the EV PDF of such processes also converge to the IID equilibrium distribution with $N=T/\Delta$. However, for potentials which increase slower than linearly, we find the opposite, namely that for large enough $T$, the EV statistics converge to those of the continuously sampled process of duration $T$.

The rest of this paper is organized as follows. In Sec.~\ref{section: cont ou} we review Majumbar, et al.'s result. Section \ref{section: disc ou} contains the main results of the paper, where subsection \ref{subsection: eigen} is the starting point of our discrete OU calculations. In subsections \ref{subsection: large z} and \ref{subsection: mu to 1} we discuss the large-$z$ asymptotics of the process and extend the continuous-sampling limit to sub-leading order, respectively, revealing in the process the scaling regime of the crossover in the large-$z$, frequently sampled limit. In Sec.~\ref{section: qualitative} we provide a qualitative argument for the origin of our findings, and in Sec.~\ref{section: general} we discuss other Langevin processes with stable equilibrium distributions, using a generalization of Majumdar, et al.'s method. We summarize our results in Sec.~\ref{section: summary}.

\section{The continuous OU process}
\label{section: cont ou}

Consider a stochastic process given as a time sequence $x(t)$, with $0\le t\le T$ and $T>0$ is the total measurement time. In the OU case, $x(t)$ evolves in time according to the Langevin equation
\begin{equation}
    \frac{\dd}{\dd t}x(t) = -\frac{1}{\tau}x(t) + \sqrt{2D}\eta(t) ,
\end{equation}
where $\tau$, $D$, and $\eta(t)$ are the relaxation time, the diffusion coefficient, and the standard Gaussian white noise, respectively, with the latter obeying $\langle\eta(t)\rangle=0$ and $\langle\eta(t)\eta(t')\rangle = \delta(t-t')$. One can always rescale the time, location, and noise as $\tilde{t} \equiv t/\tau$, $\tilde{x}(\tilde{t}) \equiv x(t)/\sqrt{D\tau}$, and $\tilde{\eta}(\tilde{t}) \equiv \sqrt{\tau}\eta(t)$, such that all quantities are dimensionless. In the foregoing we suppress the tilde notation, leading to the Langevin equation
\begin{equation}
\label{equation: langevin}
	\frac{\dd x}{\dd t} = -x(t)+\sqrt{2}\eta(t) ,
\end{equation}
i.e. all times are measured in units of the relaxation time. The equilibrium distribution of $x(t)$ is the time-independent solution of the Fokker-Planck equation
\begin{equation}
\label{equation: fokker planck}
	\frac{\partial P}{\partial t} = \frac{\partial^2P}{\partial x^2} + \frac{\partial}{\partial x} \left(xP\right) ,
\end{equation}
with vanishing boundary conditions of $P_{\rm eq}(x\to\pm\infty)=0$. The result is the zero-mean and unit-variance Gaussian, $P_{\rm eq}(x)=\phi(x)$, where
\begin{equation}
\label{equation: gaussian pdf}
    \phi(x) \equiv \frac{e^{-x^2/2}}{\sqrt{2\pi}} .
\end{equation}

Let us define the maximum of the aforementioned OU process as $z\equiv\max_{0\le t\le T}[x(t)]$. In Ref.~\cite{Majumdar}, Majumdar, et al. showed that an EV distribution for this maximum can be obtained using an eigenvalue expansion of the PDF $P_{\rm c}(x,t|z)$, which denotes the probability for a particle described by $x(t)$ to arrive at $x$ at time $t$, while always staying below the value $z$. The PDF $P_{\rm c}(x,t|z)$ obeys the Fokker-Planck equation Eq.~(\ref{equation: fokker planck}), with initial and boundary conditions of $P_{\rm c}(x,0|z)=\delta(x)$ and $P_{\rm c}(x\to-\infty,t|z)=P_{\rm c}(z,t|z)=0$, where $\delta(\cdot)$ is Dirac's delta function. A solution of Eq.~(\ref{equation: fokker planck}) was found using separation of variables to be
\begin{equation}
\label{equation: eigenvalue expansion}
	P_{\rm c}(x,t|z) = \sum_{\lambda_{\rm c}} A_{\lambda_{\rm c}}(z) \text{D}_{\lambda_{\rm c}(z)}(-x) \exp\left[-\lambda_{\rm c}(z) t -\frac{x^2}{4}\right] ,
\end{equation}
where $\text{D}_{\lambda}(\cdot)$ is the parabolic cylinder function. The set of eigenvalues $\{\lambda_{\rm c}(z)\}$ is obtained from the boundary condition at $x=z$ as the roots of
\begin{equation}
\label{equation: boundary condition continuous}
	\text{D}_{\lambda_{\rm c}(z)}(-z) = 0 .
\end{equation}
Then, at time $t=T$, the EV CDF is given by $F^{\rm c}_T(z) = \int_{-\infty}^z \dd x \, P_{\rm c}(x,T|z)$.

For large $T$, the smallest eigenvalue, which we denote by $\lambda_*$, dominates, and the typical values of $z$ are large, of order $\sim{\cal O}[\sqrt{\ln(T)}]$ \cite{Fisher,Gyorgyi,Zarfaty}. Equation~(\ref{equation: boundary condition continuous}) yields the asymptotics of $\lambda_*(z)$ for large $z$ \cite{MajumdarPC},
\begin{equation}
\label{equation: lambda g asymptotics}
	\lambda_*(z) \simeq \frac{z}{\sqrt{2\pi}} e^{-z^2/2} .
\end{equation}
The fact that $\lambda_*(z\to\infty) \to 0$ is to be expected, since in this limit the boundary conditions of $P_{\rm c}$ reduce to those of $P_{\rm eq}$, and one obtains the equilibrium density associated with a zero eigenvalue. Furthermore, taking $\lambda_* \ll 1$ gives $A_{\lambda_*} \simeq 1/\sqrt{2\pi}$ and $\text{D}_{\lambda_*}(-x) \simeq \exp(-x^2/4)$, both with exponentially small corrections, and so the EV CDF in the limit of large $T$ is
\begin{equation}
\label{equation: continuous cdf long time}
	F^{\rm c}_T(z) \simeq e^{-\lambda_*(z)T} .
\end{equation}
This large-$T$ behavior for the continuously-sampled problem is parallel to the shape of an IID EV CDF, for which $F^{\rm I}_N(z) = [F^{\rm I}(z)]^N$. In this comparison, the role of $N$ is being played by the dimensionless overall time duration $T$ (recall that all times are measured in units of $\tau$), whereas $F^{\rm c}(z) = \exp[-\lambda_*(z)]$ takes the role of $F^{\rm I}(z)$. As we saw, the equilibrium distribution of the OU process is a Gaussian, hence the latter IID case satisfies $F^{\rm I}(z) = \Phi(z)$, where
\begin{equation}
\label{equation: gaussian cdf}
    \Phi(z) \equiv \frac{1}{2}\left[1+\erf\left(\frac{z}{\sqrt{2}}\right)\right]
\end{equation}
is the standard Gaussian CDF, and $\erf(\cdot)$ is the error function. While both $F^{\rm c}(z)$ and $F^{\rm I}(z)$ approach unity as $z\to\infty$, for large $z$ we have for the former
\begin{equation}
    1-F^{\rm c}(z) \simeq \frac{z}{\sqrt{2\pi}} e^{-z^2/2} ,
\end{equation}
but for the latter we have
\begin{equation}
    1-F^{\rm I}(z) \simeq \frac{z^{-1}}{\sqrt{2\pi}}e^{-z^2/2} ,
\end{equation}
which is a factor of $z^2$ smaller.

Here we can see that the EV statistics of this case do not reproduce those of the equilibrium IID variables. Rather, they recapitulate the statistics for IID variables with an asymptotic PDF proportional to $z^2\exp(-z^2/2)$. As mentioned in the introduction, this is counter-intuitive, since for large $T$ we naively expect that the time series contains of order $T$ independent samples, drawn from the equilibrium Gaussian distribution. For this IID case, the mode of the EV distribution (the shift $b_N$ in the Gumbel variable $z_{\rm r}$ defined in the introduction) is approximately given by \cite{Fisher}
\begin{equation}
\label{equation: bn def}
    1-\Phi(b_N) \simeq \frac{e^{-b_N^2/2}}{\sqrt{2\pi}b_N} = \frac{1}{N} .
\end{equation}
Similarly setting $b_T$ to be the mode of the continuous EV distribution so that $T\lambda_*(b_T) \equiv 1$, we find for $N_{\rm eff}$, i.e. the $N$ for which the IID EV distribution has a scale of $b_T$, that
\begin{equation}
	N_{\rm eff}(T) \equiv \sqrt{2\pi} b_T \exp\left(\frac{b_T^2}{2}\right) = b_T^2 T .
\end{equation}
Hence, we see that $N_{\rm eff}$ {\em is not} proportional to $T$, but rather $N_{\rm eff}/T$ diverges, due to the factor $b_T^2$ which scales as $\ln(T)$. Next, we discuss the case of the discretely-sampled OU model, showing how the EV statistics is dramatically modified by this description. Our main point is that even if the sampling time is far less than the correlation time $\tau$, for large enough $T$ the results converge to those of the IID Gaussian, rather than to that of the continuously-sampled process.
\begin{figure}
	\includegraphics[width=1.0\textwidth]{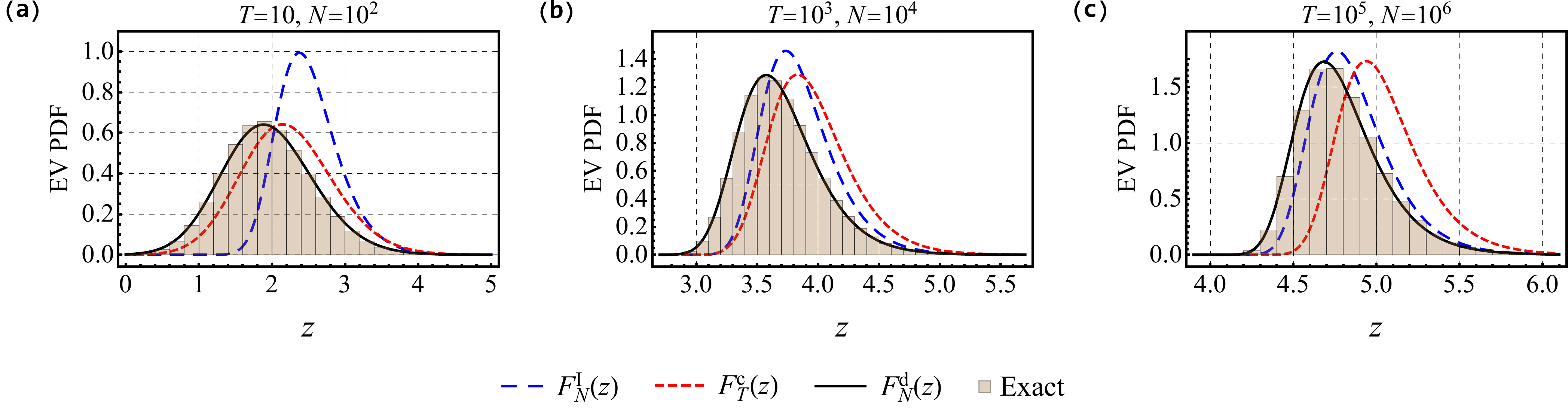}
	\caption{(Color online) {\bf The EV PDF of the discretely-sampled OU process} $x(t)$ evolving via the Langevin equation, Eq.~(\ref{equation: langevin}). This PDF demonstrates a crossover with increasing measurement time $T$ from the continuous-sampling limit to an IID behavior. We set $\Delta=0.1$ and considered three times: (a) $T=10$, (b) $T=10^3$, and (c) $T=10^5$. For each case, we sampled $10^6$ maxima using Eq.~(\ref{equation: stochastic map}), and compared the resulting histograms with the continuous-sampling and IID predictions, $F^{\rm c}_T(z)$ [via Eqs.~(\ref{equation: boundary condition continuous}) and (\ref{equation: continuous cdf long time})] and $F^{\rm I}_N(z)=\Phi^N(z)$ [via Eq.~(\ref{equation: gaussian cdf})], respectively, with $N=T/\Delta$. We also show the exact distribution Eq.~(\ref{equation: discrete cdf}), obtained from numerical evaluations of the eigenvalue equation, Eq.~(\ref{equation: eigensystem}). For short times, we find that the exact results are closer to the continuous-sampling limit, see panel (a). This begins to change significantly when $T$ is of order a crossover scale $T_{\rm cross}(\Delta)$, to be discussed in Sec.~\ref{section: qualitative}, see panel (b). For $T \gg T_{\rm cross}$, the exact values start to converge to the IID limit, see panel (c). The initial condition for all samplings was $x=0$. We used $A(z)=1$ in Eq.~(\ref{equation: discrete cdf}) when plotting $F^{\rm d}_N(z)$, see main text.}
\label{figure: ou samples}
\end{figure}

\section{The discretely sampled OU process}
\label{section: disc ou}

To answer the questions raised in the introduction, we turn to consider the problem of sampling the OU process at a finite time interval $\Delta$. Using the exact solution of the OU process Eq.~(\ref{equation: langevin}), we can write a stochastic map which directly yields the discretely-sampled values $x_n \equiv x(n\Delta)$ as
\begin{equation}
\label{equation: stochastic map}
	x_{n+1}= \mu x_n + \sqrt{1-\mu^2}\eta_n ,
\end{equation}
where $\eta_n$ is a standard Gaussian deviate. This definition of $x_n$ is a generalization of the $x^{\rm d}_n$ defined in the introduction, where $\tau$ was replaced by $\Delta$. This also calls for redefining the number of IID RVs that was introduced in the comparison above as $N \equiv T/\Delta$, such that $0\le n\le N$ (with $\Delta=1$ yielding the previous definitions). It is easy to see that Eq.~(\ref{equation: stochastic map}) results from Eq.~(\ref{equation: langevin}), as the latter has the solution
\begin{equation}
	x(t+\Delta)=e^{-\Delta}x(t) + \sqrt{2} \int_0^\Delta \dd t' \, e^{t'-\Delta} \eta(t') .
\end{equation}
The integral term is a Gaussian random variable with mean $0$ and standard deviation $(1-e^{-2\Delta})^{1/2}$, thus our discrete mapping coincides with Eq.~(\ref{equation: langevin}) if $\mu = e^{-\Delta}$. From here on, $z$ refers to the discrete-sampling EV, $z\equiv\max_{0\le n\le N}(x_n)$, and as pointed out in the introduction, this maximum is always smaller than or equal to that of the continuously-sampled process.

\subsection{The eigenvalue equation}
\label{subsection: eigen}

If we start out with some initial distribution of $x$, $P_0(x)$ [e.g. a localized initial condition at the origin means that $P_0(x)=\delta(x)$], then the distribution of $x$ after $n$ iterations of Eq.~(\ref{equation: stochastic map}), $P_n(x)$, satisfies
\begin{equation}
	P_n(x) = \int_{-\infty}^{\infty} \dd x' \, P_{n-1}(x') \frac{1}{\sqrt{2\pi(1-\mu^2)}} \exp\left[-\frac{(x-\mu x')^2}{2(1-\mu^2)}\right] ,
\end{equation}
since the noise $\eta_n$ in Eq.~(\ref{equation: stochastic map}) has a standard Gaussian distribution. Clearly, the EV is smaller than $z$ if and only if all the $x_n$s are less than $z$. The probability of this event for $x_0$ alone is trivially given by $\text{Prob}(x_0\le z) = \int_{-\infty}^z \dd x' \, P_0(x';z)$, where $P_0(x;z)=\theta(z-x)P_0(x)$ is the (truncated) initial condition, and $\theta(\cdot)$ is the Heaviside step function. For $N=1$, we have $\text{Prob}[\max(x_0,x_1)\le z] = \int_{-\infty}^z \dd x \int_{-\infty}^z \dd x' \, \text{Prob}(x_0=x' \wedge x_1=x)$. Using Bayes' theorem and Eq.~(\ref{equation: stochastic map}) yields $\text{Prob}(x_0=x' \wedge x_1=x) = \theta(z-x) P_0(x';z) \exp[-(x-\mu x')^2/2]/\sqrt{2\pi(1-\mu^2)}$. Hence, we conclude that the EV CDF for this discrete case with $N>1$ is given by $F^{\rm d}_N(z) \equiv \int_{-\infty}^z\dd x\,P_N(x;z)$, where $P_n(x;z)$ satisfies the recurrence relation
\begin{equation}
\label{equation: iterations}
	P_n(x;z) = \theta(z-x) \int_{-\infty}^z \dd x' \, P_{n-1}(x';z) \frac{1}{\sqrt{2\pi(1-\mu^2)}} \exp\left[-\frac{(x-\mu x')^2}{2(1-\mu^2)}\right] .
\end{equation}
Equation~(\ref{equation: iterations}) is a linear map from $P_{n-1}$ to $P_n$, and so it can be solved via an eigenvalue expansion,
\begin{equation}
    P_n(x;z)=\sum_{\Lambda} A_{\Lambda}(z) \Lambda^n(z) P_{\Lambda}(x;z) ,
\end{equation}
similar to Eq.~(\ref{equation: eigenvalue expansion}), with an eigenfunction equation
\begin{equation}
	\Lambda(z)P_\Lambda(x;z) = \theta(z-x) \int_{-\infty}^z \dd x' \, P_\Lambda(x';z) \frac{1}{\sqrt{2\pi(1-\mu^2)}} \exp\left[-\frac{(x-\mu x')^2}{2(1-\mu^2)}\right] .
\end{equation}
Due to the $\theta$-function cutoff at $x=z$, probability is lost in each iteration, and so all the eigenvalues are smaller than unity. Thus, for large $n$, this solution is dominated by the eigenvalue of largest magnitude $\Lambda_*(z)$ and its corresponding eigenfunction $P_*(x;z)$, which are the central objects of our investigation, satisfying
\begin{equation}
\label{equation: eigensystem}
	\Lambda_*(z)P_*(x;z) = \theta(z-x) \int_{-\infty}^z \dd x' \, P_*(x';z) \frac{1}{\sqrt{2\pi(1-\mu^2)}} \exp\left[-\frac{(x-\mu x')^2}{2(1-\mu^2)}\right] .
\end{equation}
The shape of $P_n(x;z)$ converges for large $n$ to $P_n(x;z) \simeq A_{\Lambda_*}(z)\Lambda_*^n(z)P_*(x;z)$, and therefore the EV CDF obeys
\begin{equation}
\label{equation: discrete cdf}
	F^{\rm d}_N(z) \simeq A(z) e^{-N\ln[1/\Lambda_*(z)]} , \quad A(z) \equiv A_{\Lambda_*}(z) \int_{-\infty}^z \dd x \, P_*(x;z) \mathop{\to}_{z\to\infty} 1 ,
\end{equation}
where the large-$z$ limit of $A(z)$ directly follows from the fact that any CDF $F(x)$ with infinite support $x\in(-\infty,\infty)$ obeys $F(x\to\infty)\to 1$. Note that the initial condition dependency enters via the prefactor $A_{\Lambda}(z)$, but the large-$z$ limit is independent of the initial condition.

Numerically, it is straightforward to find $\Lambda_*(z)$, and the numerical scheme is described in Appendix~\ref{appendix: numerics}. Figure~\ref{figure: ou samples} shows the EV PDF of the discretely-sampled OU process with a sampling interval $\Delta=0.1$, for three measurement times $T=10,10^3,10^5$. The PDFs derived from Eqs.~(\ref{equation: eigensystem}) and (\ref{equation: discrete cdf}) excellently match the sampled histograms, while both the continuous-sampling and IID limits, given by Eqs.~(\ref{equation: continuous cdf long time}) and (\ref{equation: gaussian cdf}) respectively (with $F^{\rm I}_N=\Phi^N$), fail.
\begin{figure}
	\includegraphics[width=0.75\textwidth]{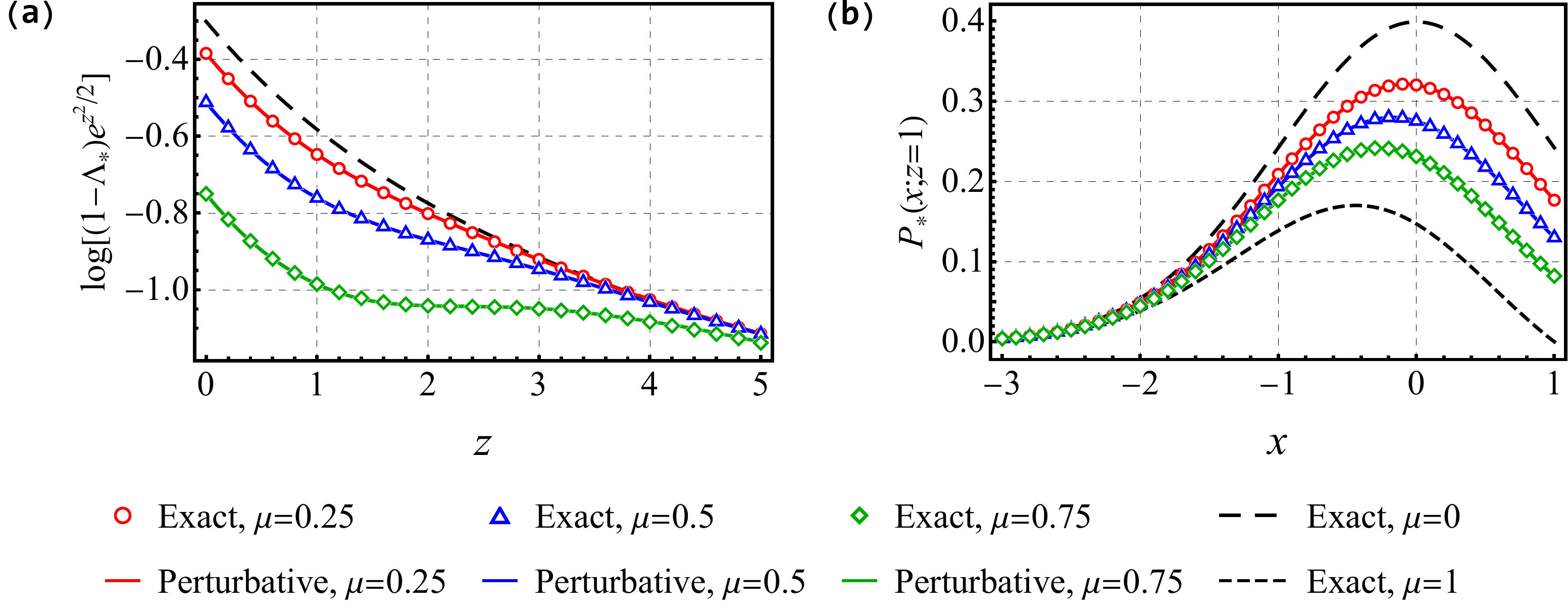}
	\caption{(Color online) {\bf The solution of the discrete eigenvalue equation:} (a) The scaled eigenvalue $(1-\Lambda_*)\exp(z^2/2)$ for $\mu=0.25$, $0.5$, and $0.75$, from direct numerical evaluations of the eigenvalue equation, Eq.~(\ref{equation: eigensystem}), as well as from a $10$th-order expansion in $\mu$. Also shown is the exact result for the IID case, $\mu=0$, for which $\Lambda_* = \lambda_0 = \Phi$. Notice that all three finite-$\mu$ curves merge for large $z$ with the IID curve. (b) The eigenfunction $P_*(x;z)$ for $z=1$, normalized such that $P_*(-5;1)=\phi(-5)$. A $20$th-order expansion for the three values of $\mu$ excellently matches with direct numerical evaluations of the eigenvalue equation, Eq.~(\ref{equation: eigensystem}). We also add the exact results for the IID and continuous-sampling limits, given by Eqs.~(\ref{equation: gaussian pdf}) and (\ref{equation: continuous x pdf long time}), respectively.}
\label{figure: lambda and p for small mu}
\end{figure}

The long-time behavior of the continuous-sampling limit can be retrieved from Eq.~(\ref{equation: eigensystem}) by taking $\Delta\to 0$, leading to $\mu \simeq 1-\Delta$ and $\Lambda_* \simeq 1-\lambda_*\Delta$. Changing variables to $\chi=(x-\mu x')/\sqrt{1-\mu^2}$ in the integration and expanding for $\Delta\to 0$ gives
\begin{equation}
\label{equation: mu to 1 evidence}
	P_*(x;z) \simeq P_*(x;z) + \Delta \left\{
	    \begin{aligned}
	        &\left[1+\lambda_*(z)\right]P_*(x;z)+x\frac{\dd}{\dd x}P_*(x;z)+\frac{\dd^2}{\dd x^2} P_*(x;z) & x<z \\
	        &-\frac{1}{2\Delta} P_*(z;z) & x=z
	    \end{aligned}
	\right. .
\end{equation}
Hence, to leading order, $P_*(x;z)$ satisfies the differential equation
\begin{equation}
	\frac{\dd^2}{\dd x^2} P_*(x;z) + \frac{\dd}{\dd x} \left[P_*(x;z)x\vphantom{\frac{1}{1}}\right] = -\lambda_*(z) P_*(x;z) ,
\end{equation}
with a boundary condition at $x=z$ of $P_*(z;z)=0$, giving the solution
\begin{equation}
\label{equation: continuous x pdf long time}
	P_*(x;z) \propto \exp\left(-\frac{x^2}{4}\right)\text{D}_{\lambda_*(z)}(-x) , \quad \text{D}_{\lambda_*(z)}(-z) = 0 ,
\end{equation}
exactly as in Ref.~\cite{Majumdar} and Sec.~\ref{section: cont ou}.

We can also solve Eq.~(\ref{equation: eigensystem}) in the large $\Delta$ (i.e. small $\mu$) limit via perturbation theory. We use the following ansatz,
\begin{equation}
    \Lambda_*(z) = \sum_{n=0}^{\infty}\lambda_n(z)\mu^n , \quad P_*(x;z) = \theta(z-x)\phi(x) \left[ 1 + \sum_{n=1}^{\infty}\mu^n \sum_{m=1}^n p_{nm}(z)x^m \right] ,
\end{equation}
where as before, $\phi(x)$ is the standard Gaussian distribution. To zeroth order in $\mu$ we get $\lambda_0(z) = \Phi(z)$, which is to be expected since for $\mu\to 0$ the $x_n$s are IID Gaussian variables. To next order,
\begin{equation}
	p_{11}(z) = -\sqrt{\frac{2}{\pi}} \frac{e^{-z^2/2}}{1 + \erf(z/\sqrt{2})} , \quad \lambda_1(z) = \frac{e^{-z^2}}{\pi[1 + \erf(z/\sqrt{2})]} .
\end{equation}
It is straightforward to continue this calculation, which we have carried out to order $\mu^{20}$. The results agree extremely well with the direct numerical evaluations of Eq.~(\ref{equation: eigensystem}). Figure~\ref{figure: lambda and p for small mu}(a) shows the scaled eigenvalue $(1-\Lambda_*)\exp(z^2/2)$ for $\mu=0.25$, $0.5$, and $0.75$, along with the IID result, i.e. $\mu=0$, given by Eq.~(\ref{equation: gaussian cdf}). The eigenfunction $P_*(x,z)$ is seen in Fig.~\ref{figure: lambda and p for small mu}(b) for $z=1$ and the same $\mu$s, together with the IID and continuous-sampling limits derived above, given by Eqs.~(\ref{equation: gaussian pdf}) and (\ref{equation: continuous x pdf long time}), respectively.
\begin{figure}
	\includegraphics[width=0.4\textwidth]{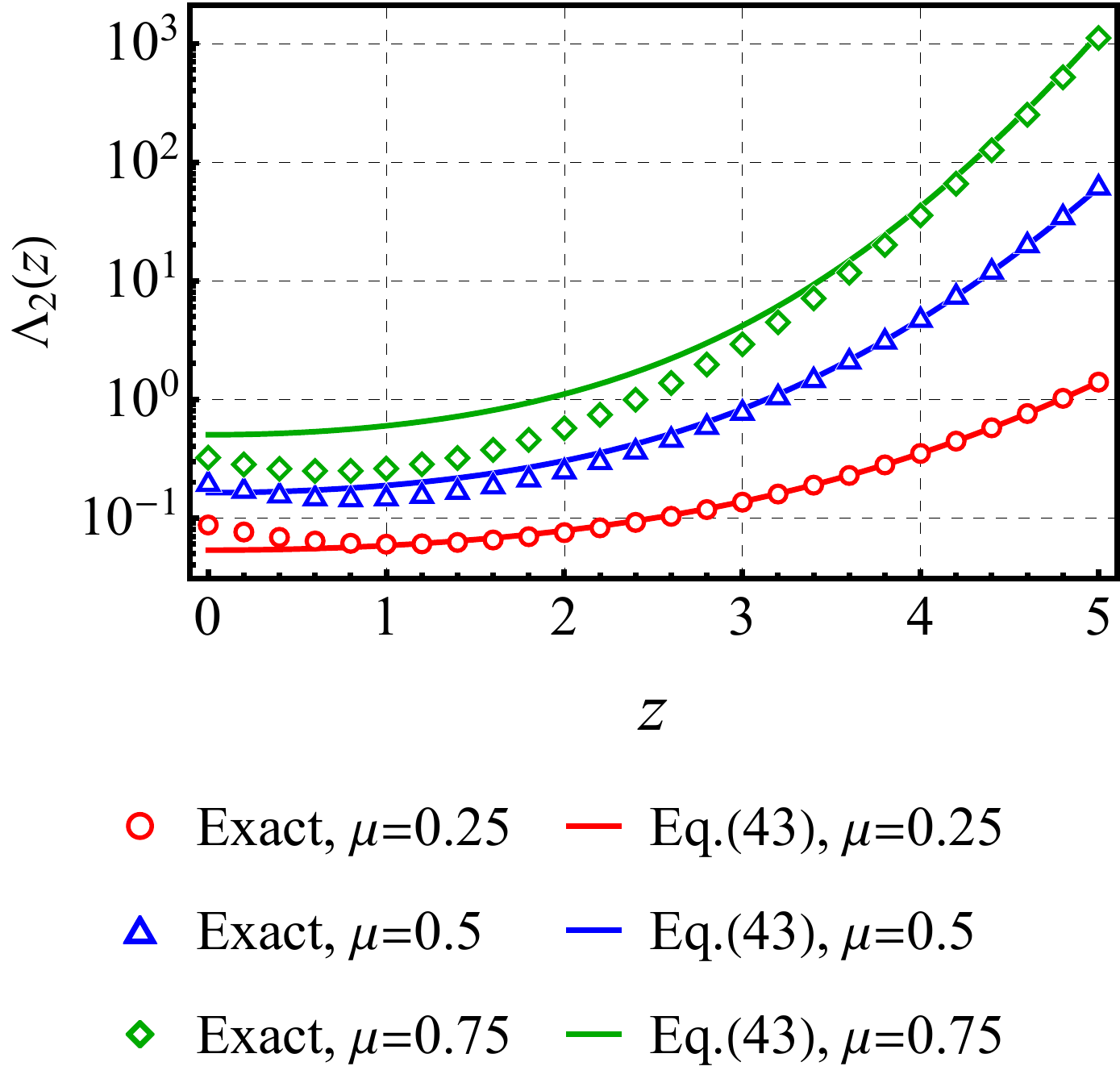}
	\caption{(Color online) {\bf The second correction to $\Lambda_*$ in the large-$z$ approximation}, $\Lambda_2 \simeq [\Lambda_*-1-\exp(-z^2/2)\Lambda_1]\exp(z^2)$, where $\Lambda_1$ is given by Eq.~(\ref{equation: large z first order}). Seen is the numerical data of Fig.~\ref{figure: lambda and p for small mu}(a), together with our prediction given by Eq.~(\ref{equation: lambda 2}), truncated at $n=25$. The two agree well for large enough $z$s. As $\mu\to 1$, one requires larger $z$s for the correction to be accurately described by Eq.~(\ref{equation: lambda 2}).}
\label{figure: lambda 2}
\end{figure}

The most striking aspect of Fig.~\ref{figure: lambda and p for small mu}(a) is that even though increasing $\mu$ increases $\Lambda_*$, all three curves appear to merge with the IID curve at large $z$. This seems to indicate that for large $N$, the discretely-sampled EV statistics converge to the IID EV statistics, since as mentioned below Eq.~(\ref{equation: boundary condition continuous}), the large-$T$ (or alternatively, large-$N$) EV behavior is governed by the large-$z$ behavior of the eigenvalue. This observation is strengthened by examining the large-$z$ asymptotics of the perturbative result, where $p_{11} \propto \exp(-z^2/2)$ and $\lambda_1 \propto \exp(-z^2)$. Therefore, $\lambda_1$ makes only an {\em exponentially small} contribution to $\ln(\Lambda_*)$, swamped by $\ln(\lambda_0) \propto \exp(-z^2/2)$. Similarly, $p_{ij}\propto \exp(-z^2/2)$ and $\lambda_i\propto \exp(-z^2)$ for the first twenty orders we have calculated. If this continues to hold true also for all the higher-order $\mu$ corrections, this gives us our essential finding. Namely, that the EV statistics of the discretely-sampled OU process has exactly the same large-$N$ (or equivalently, large-$T$) behavior as uncorrelated Gaussian variables, for any $0\le\mu<1$. Returning to Fig.~\ref{figure: ou samples}, we direct the reader's attention to the predictions of the continuous-sampling and IID approximations, for which $\Delta \to 0$ and $\mu \to 0$, respectively. As mentioned, these have CDFs of $F^{\rm c}_T(z)=\exp[-\lambda_*(z)T]$ and $F^{\rm I}_N(z)=\Phi^N(z)$ respectively, with $N=T/\Delta$. We see that while at short times the exact results are close to the continuous-sampling curve, see panel (a), increasing $T$ in panels (b) and (c) makes the exact values approach the IID limit curve. In the next section we prove this conjecture.

\subsection{The large-$z$ asymptotics}
\label{subsection: large z}

Based on the above arguments, for large-$z$ we write
\begin{equation}
	P_*(x;z) \simeq \phi(x)\left[1 + e^{-z^2/2}{\cal P}_1(x;z) + e^{-z^2}{\cal P}_2(x;z)\right] , \quad \Lambda_*(z) \simeq 1 + e^{-z^2/2}\Lambda_1(z) + e^{-z^2}\Lambda_2(z) .
\end{equation}
These expansions are to be understood in the context of a fixed $\mu<1$. During the following subsections we suppress the notation of the Heaviside step function. Plugging the above expansion into Eq.~(\ref{equation: eigensystem}), we get to first order
\begin{equation}
    \phi(x)\left[1 + e^{-z^2/2}{\cal P}_1(x;z) + e^{-z^2/2}\Lambda_1(z)\right] = \int_{-\infty}^z \dd x' \, \phi(x') \frac{1+e^{-z^2/2}{\cal P}_1(x';z)}{\sqrt{2\pi(1-\mu^2)}}\exp\left[-\frac{(x-\mu x')^2}{2(1-\mu^2)}\right] .
\end{equation}
The zeroth-order equation is satisfied since
\begin{equation}
    \int_{-\infty}^z \dd x' \, \phi(x')\frac{1}{\sqrt{2\pi(1-\mu^2)}}\exp\left[-\frac{(x-\mu x')^2}{2(1-\mu^2)}\right]=\phi(x)\left\{1-\frac{1}{2}\erfc\left[\frac{z-x\mu}{\sqrt{2(1-\mu^2)}}\right]\right\} ,
\end{equation}
where $\erfc(\cdot)$ is the complementary error function.
\begin{figure}
	\includegraphics[width=0.4\textwidth]{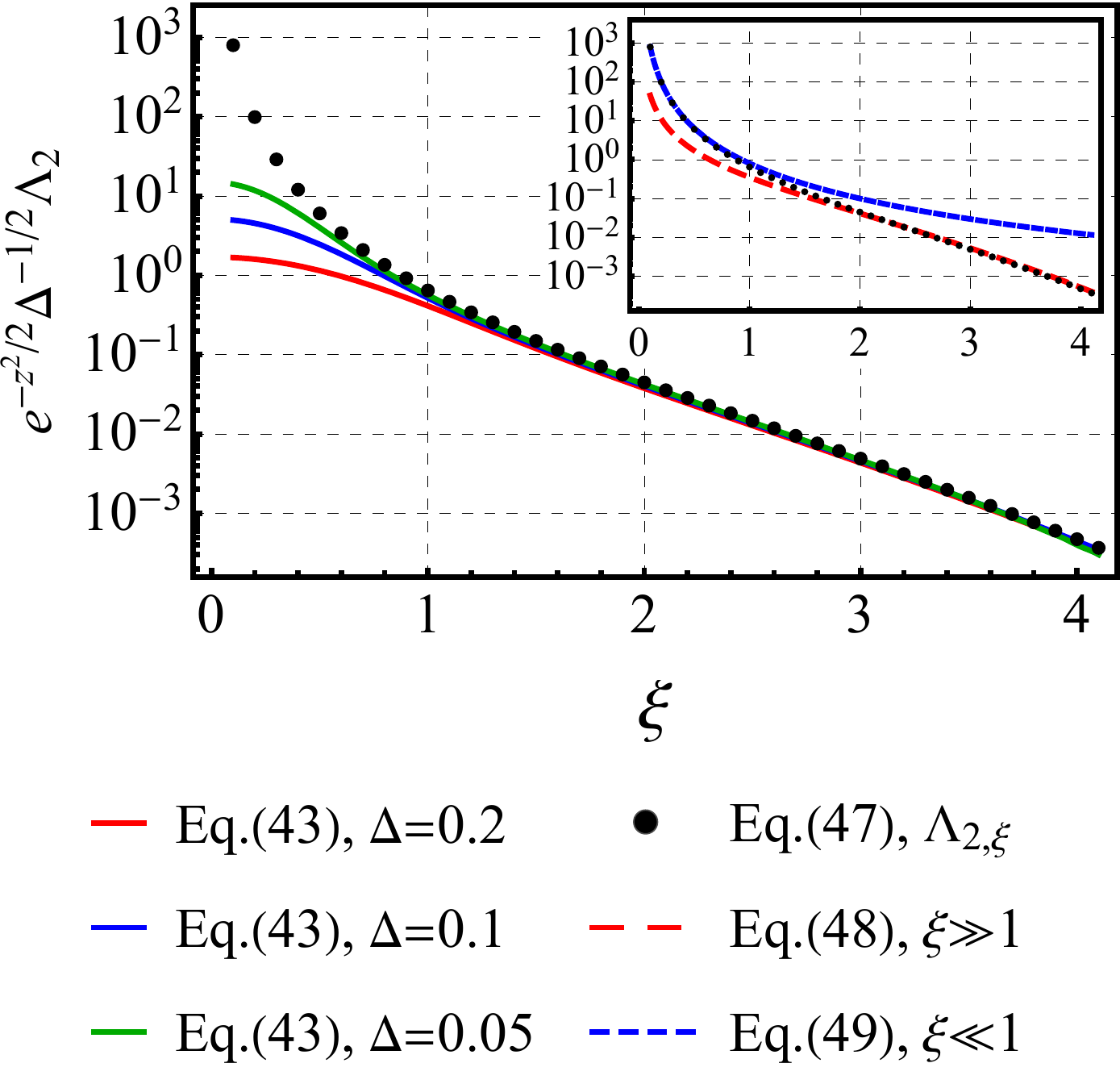}
	\caption{(Color online) {\bf The large $z$, small $\Delta$ crossover regime:} The rescaled second correction to $\Lambda_*$, $\exp(-z^2/2)\Delta^{-1/2}\Lambda_2$, plotted versus the crossover variable $\xi=\sqrt{\Delta}z$. Seen are the exact values calculated from Eq.~(\ref{equation: lambda 2}), with the summation truncated at $n=100$, for three values of $\Delta$. As $\Delta\to 0$, a data collapse of $\Lambda_2$ onto the limiting $\Lambda_{2,\xi}$, given by Eq.~(\ref{equation: lambda xi}), can be seen. $\Lambda_{2,\xi}$, together with its large- and small-$\xi$ asymptotics as given by Eqs.~(\ref{equation: lambda 2 large xi}) and (\ref{equation: lambda 2 small xi}), respectively, are given in the inset.}
\label{figure: lambda xi}
\end{figure}

Using the following expansion \cite{Hermite1} of the kernel function of Eq.~(\ref{equation: eigensystem}), which holds for $0\le\mu<1$,
\begin{equation}
\label{equation: hermite expansion 1}
	\frac{1}{\sqrt{2\pi(1-\mu^2)}}\exp\left[-\frac{(x-x'\mu)^2}{2(1-\mu^2)}\right] = \phi(x) \sum_{n=0}^{\infty} \frac{\mu^n}{n!}\He_n(x)\He_n(x') ,
\end{equation}
where $\He_n(\cdot)$ is the $n$th probabilist's Hermite polynomial, we obtain to first order
\begin{equation}
\label{equation: dropped correction}
	{\cal P}_1(x;z) + \Lambda_1(z) + \frac{1}{2}e^{z^2/2}\erfc\left[\frac{z-x\mu}{\sqrt{2(1-\mu^2)}} \right] - \int_{-\infty}^{\infty}\dd x'\,\phi(x'){\cal P}_1(x';z)\sum_{n=0}^{\infty} \frac{\mu^n}{n!}\He_n(x)\He_n(x') = 0 .
\end{equation}
Note that we have extended the integral's boundary to infinity, dropping a higher-order correction to be accounted for during the second-order calculation. Exploiting another expansion \cite{Hermite2} similar to the one above,
\begin{equation}
\label{equation: hermite expansion 2}
	\erfc\left[\frac{z-x\mu}{\sqrt{2(1-\mu^2)}}\right] = \erfc\left(\frac{z}{\sqrt{2}}\right) + 2\phi(z) \sum_{n=1}^{\infty} \frac{\mu^n}{n!}\He_n(x)\He_{n-1}(z) ,
\end{equation}
together with expressing the first functional correction as a sum over Hermite polynomials in $x$,
\begin{equation}
	{\cal P}_1(x;z) = \sum_{n=0}^{\infty} c_n(z) \He_n(x) ,
\end{equation}
and using their orthogonality condition (where $\delta_{n,m}$ is the Kronecker delta),
\begin{equation}
\label{equation: hermite orthogonality}
    \int_{-\infty}^{\infty} \dd x' \phi(x')\He_{n}(x')\He_m(x') = \delta_{n,m} n! ,
\end{equation}
we get for the first-order expansion
\begin{equation}
	\left[c_0(z)\left(1-\mu^0\right)+\Lambda_1(z)+\frac{1}{2}e^{z^2/2}\erfc\left(\frac{z}{\sqrt{2}}\right)\right]\He_0(x) + \sum_{n=1}^{\infty}\left[c_n(z)\left(1-\mu^n\right)+ \frac{\mu^n/n!}{\sqrt{2\pi}}\He_{n-1}(z)\right]\He_n(x) = 0 .
\end{equation}
Thus, we obtain
\begin{equation}
\label{equation: large z first order}
	\Lambda_1(z) = -\frac{1}{2}e^{z^2/2}\erfc\left(\frac{z}{\sqrt{2}}\right) , \quad c_n(z) = -\frac{\mu^n/n!}{\sqrt{2\pi}}\frac{\He_{n-1}(z)}{1-\mu^n} , \quad n>0 .
\end{equation}
The value of $c_0(z)$ can be found from the condition ${\cal P}_1(0;z)=0$, since an $x$-independent addition to ${\cal P}_1$ is just a change of normalization. This yields
\begin{equation}
	c_0(z) = -\sum_{n=1}^{\infty} c_n(z) \He_n(0) = -\sum_{n=1}^{\infty} \frac{\sqrt{\pi}2^{n/2}c_n(z)}{\Gamma[(1-n)/2]} ,
\end{equation}
where $\Gamma(\cdot)$ is the gamma function. We can also use Eq.~(\ref{equation: hermite expansion 2}) to express ${\cal P}_1(x;z)$ as
\begin{equation}
	{\cal P}_1(x;z) = \frac{1}{2}e^{z^2/2}\sum_{n=0}^{\infty}\left\{ \erfc\left[ \frac{z}{\sqrt{2[1-\mu^{2(n+1)}]}}\right] - \erfc\left[ \frac{z-x\mu^{n+1}}{\sqrt{2[1-\mu^{2(n+1)}]}}\right] \right\} .
\end{equation}
Note that as expected, for $\mu\to 0$ this first correction vanishes, as $\mu=0$ represents the IID case, where the distribution of the discretely-sampled process reduces to the equilibrium Gaussian.
\begin{figure}
	\includegraphics[width=0.75\textwidth]{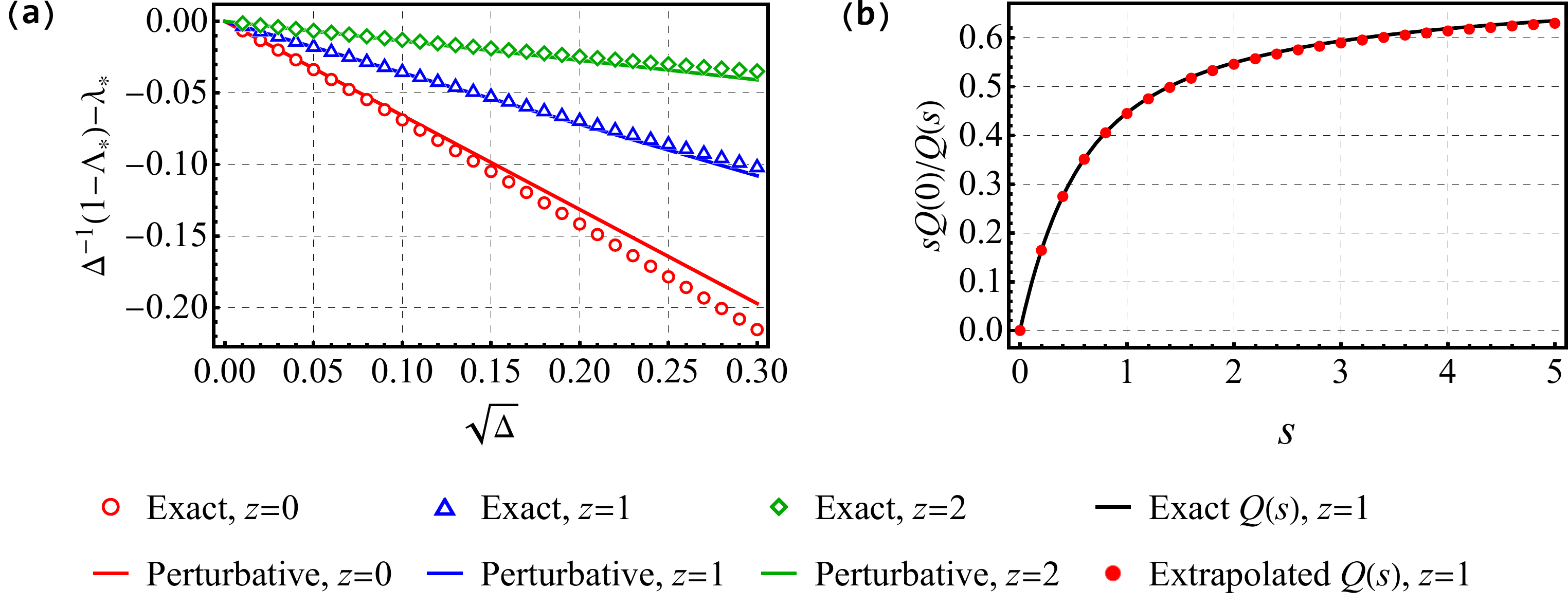}
	\caption{(Color online) {\bf The scaling of the eigenvalue equation solution in the small $\Delta$ regime:} (a) The difference between the scaled eigenvalue and the continuous-sampling limit, $\Delta^{-1}(1-\Lambda_*)-\lambda_*$, for $z=0$, $1$, and $2$, obtained from direct numerical evaluations of the eigenvalue equation, Eq.~(\ref{equation: eigensystem}), as well as the approximation for $\Delta \to 0$, $\Lambda_* \simeq 1 - \Delta (\lambda_*+\sqrt{\Delta}\lambda_{1/2})$. (b) The eigenfunction $P_*(x;z)$, normalized to unity at $x=z$ [note that $P_*(x,z)/P_*(z,z)=Q(s)/Q(0)$], for $\Delta=0$. Seen are extrapolations over $\sqrt{\Delta}$ of the numerical evaluations of Eq.~(\ref{equation: eigensystem}), together with the inner solution $Q(s)$ (see more details at the end of appendix~\ref{appendix: inner}).}
\label{figure: lambda and p for fixed z}
\end{figure}

Since we have an exact solution of the first-order equation, we can move on to the second order. We further expand Eq.~(\ref{equation: eigensystem}) to second-order, obtaining
\begin{align}
	{\cal P}_2(x;z)+{\cal P}_1(x;z)\Lambda_1(z)+\Lambda_2(z) + e^{z^2/2} &\int_z^{\infty} \dd x'\,\phi(x'){\cal P}_1(x';z) \sum_{n=0}^{\infty} \frac{\mu^n}{n!}\He_n(x)\He_n(x') \nonumber \\
	- &\int_{-\infty}^{\infty} \dd x'\,\phi(x'){\cal P}_2(x';z) \sum_{n=0}^{\infty} \frac{\mu^n}{n!}\He_n(x)\He_n(x') = 0 .
\end{align}
The first integral term is the higher-order correction that was dropped in Eq.~(\ref{equation: dropped correction}). As was done above, the boundary of the second integral term was extended to infinity (as the contribution from $z$ to infinity only enters the calculation of the third-order correction). Expressing ${\cal P}_2(x;z)$ similarly to its first-order counterpart,
\begin{equation}
	{\cal P}_2(x;z) = \sum_{n=0}^{\infty} d_n(z) \He_n(x) ,
\end{equation}
and rearranging the terms, we get
\begin{equation}
	\Lambda_2(z) = \sum_{n=1}^{\infty} \frac{\mu^n/n!}{1-\mu^n} \phi_{n,0}(z)\He_{n-1}(z) ,
\end{equation}
with \cite{Hermite3}
\begin{equation}
\label{equation: hermite expansion 3}
	\phi_{n,m}(z) \equiv \frac{e^{z^2/2}}{\sqrt{2\pi}} \int_z^{\infty}\dd x\,\phi(x) \He_n(x)\He_m(x) = \frac{1}{2\pi}\sum_{l=0}^L l!\binom{n}{l}\binom{m}{l} \He_{n+m-2l-1}(z) + \delta_{n,m}\frac{n!/2}{\sqrt{2\pi}} e^{z^2/2}\erfc\left( \frac{z}{\sqrt{2}}\right) ,
\end{equation}
where $L \equiv \min(n,m)-\delta_{n,m}$ and we used the standard convention that a summation from $0$ to $-1$ vanishes. Taking $m=0$ and $n\ge1$, we get $\phi_{n,0}(z)=(1/2\pi)\He_{n-1}(z)$, hence
\begin{equation}
\label{equation: lambda 2}
	\Lambda_2(z) = \frac{1}{2\pi} \sum_{n=1}^{\infty} \frac{\mu^n/n!}{1-\mu^n} \He_{n-1}^2(z) .
\end{equation}
Figure~\ref{figure: lambda 2} shows $\Lambda_2 \simeq [\Lambda_*-1-\exp(-z^2/2)\Lambda_1]\exp(z^2)$, where $\Lambda_1$ is given by Eq.~(\ref{equation: large z first order}), for the sampled data of Fig.~\ref{figure: lambda and p for small mu}(a). Also seen is the prediction of $\Lambda_2$, Eq.~(\ref{equation: lambda 2}), with the summation truncated at $n=25$. The closer $\mu$ is to $1$, the larger $z$ has to be for the prediction to be accurate. Next, using the identity
\begin{equation}
	\sum_{n=0}^{\infty} \frac{\nu^n}{n!} \He_n^2(z) = \frac{1}{\sqrt{1-\nu^2}} \exp\left(\frac{z^2\nu}{1+\nu}\right) ,
\end{equation}
which arises in the calculation of the density of states of the finite temperature quantum harmonic oscillator \cite{Bondarev}, we find
\begin{subequations}
    \begin{align}
    \label{equation: lambda 2 solution}
	    \Lambda_2(z) & = \frac{1}{2\pi} \sum_{n=1}^{\infty} \int_0^{\mu^n} \frac{\dd\nu}{\sqrt{1-\nu^2}} \exp\left(\frac{z^2\nu}{1+\nu}\right) \\
	    & \mathop{\sim}_{z\to\infty} \frac{(1+\mu)^2}{\sqrt{1-\mu^2}} \frac{1}{2\pi z^2} \exp\left(\frac{z^2\mu}{1+\mu}\right) .
    \end{align}
\end{subequations}
Thus, the overall large-$z$ behavior of the $\Lambda_2$ contribution to $\Lambda_*$ is dominated by the $n=1$ term of the sum and is proportional to $\exp[-z^2/(1+\mu)]$, which varies from $\exp(-z^2)$ for $\mu\to 0$ to $\exp(-z^2/2)$ as $\mu \to 1$. Thus, for any finite $\Delta$, $\Lambda_*$ is dominated by the IID $\Lambda_1$ contribution for large $z$. One can obtain an exact solution for ${\cal P}_2(x;z)$ similarly to its first-order counterpart. The resulting formula is quite cumbersome, and does not contribute to the rest of the discussion, and hence it is omitted.
\begin{figure}
	\includegraphics[width=0.75\textwidth]{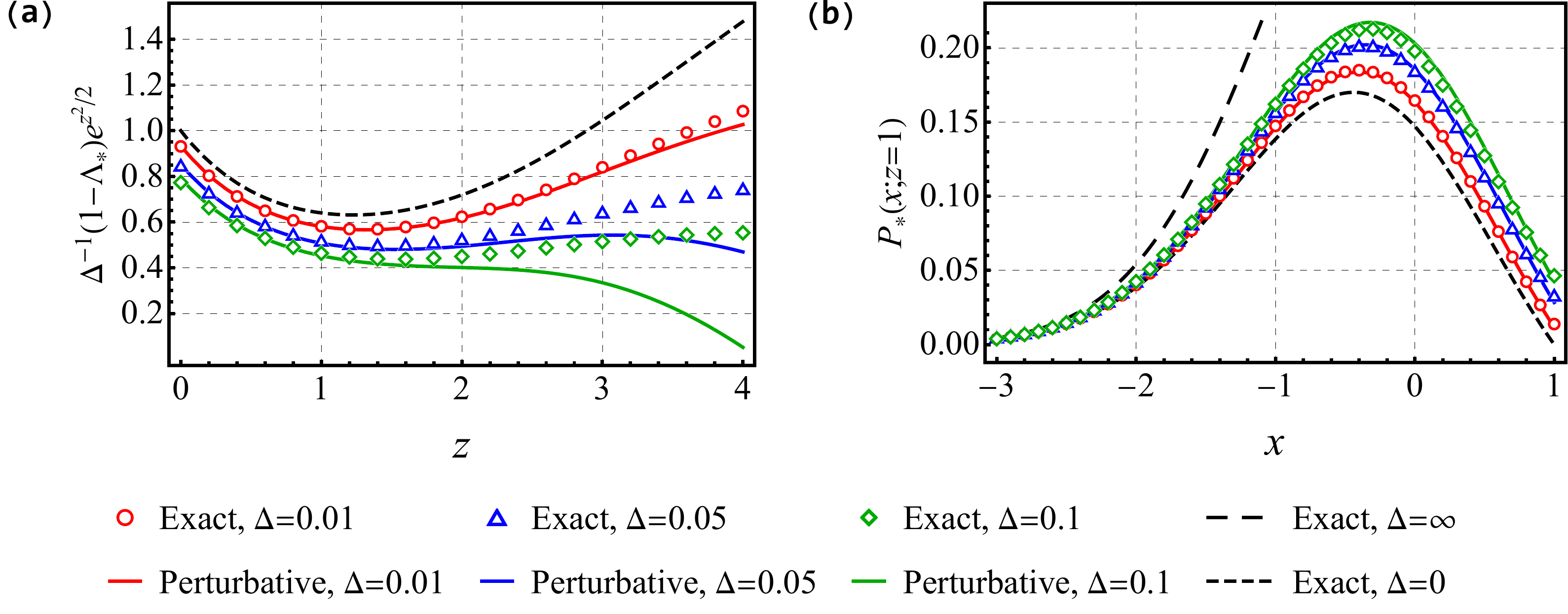}
	\caption{(Color online) {\bf The eigenvalue solution in the small $\Delta$ regime:} (a) The scaled eigenvalue $\Delta^{-1}(1-\Lambda_*)\exp(z^2/2)$ for $\Delta=0.01$, $0.05$, and $0.1$, obtained from direct numerical evaluations of the eigenvalue equation, Eq.~(\ref{equation: eigensystem}), as well as the approximation for $\Delta \to 0$, $\Lambda_* \simeq 1 - \Delta (\lambda_*+\sqrt{\Delta}\lambda_{1/2})$. We added the exact result for the continuous-sampling case, $\lambda_*\exp(z^2/2)$. (b) The eigenfunction $P_*(x;z)$ for $z=1$, normalized such that $P_*(-5,1)=\phi(-5)$. The perturbative solution of the regime $\Delta\to 0$, Eq.~(\ref{equation: outer eigenfunction}), for the three values of $\Delta$, nicely match with the numerical evaluations of Eq.~(\ref{equation: eigensystem}). We added the exact results for the IID and continuous-sampling limits, given by Eqs.~(\ref{equation: gaussian pdf}) and (\ref{equation: continuous x pdf long time}), respectively.}
\label{figure: lambda and p for large mu}
\end{figure}

Returning to Eq.~(\ref{equation: lambda 2 solution}), we see that as $\mu\to 1$, all of the terms in the sum become of the same order for $z\to \infty$. Moreover, for $\Delta\to 0$, $z\to\infty$, $\xi \equiv \sqrt{\Delta}z$ fixed, the integral in Eq.~(\ref{equation: lambda 2 solution}) has a different asymptotic limit, due to the singular behavior of the square root factor in this regime. Changing the integration variable to $\sigma \equiv 1-\nu$, we obtain
\begin{equation}
    \int_0^{\mu^n} \frac{\dd\nu}{\sqrt{1-\nu^2}}\exp\left(\frac{z^2\nu}{1+\nu}\right) \simeq \int_{n\Delta}^{\infty} \frac{\dd\sigma}{\sqrt{2\sigma}}\exp\left[\frac{z^2}{2} \left(1 - \frac{\sigma}{2}\right)\right] = \frac{\sqrt{2\pi\Delta}}{\xi}e^{z^2/2}\erfc\left(\frac{\sqrt{n}\xi}{2}\right) ,
\end{equation}
where we extended the upper boundary of the integral from $1$ to $\infty$, as the resulting correction is exponentially small. Summing over $n$ using the integral representation of $\erfc(x)=\sqrt{4/\pi}\int_x^{\infty}\dd \chi\exp(-\chi^2)$ gives
\begin{equation}
\label{equation: lambda xi}
    \Lambda_{2,\xi} \simeq \sqrt{\frac{\Delta}{2}} \frac{e^{z^2/2}}{\pi\xi} \int_{\xi}^{\infty} \dd x \, \text{Li}_{-1/2} \left(e^{-x^2/4}\right) ,
\end{equation}
where $\text{Li}_{-1/2}(\cdot)$ is the polylogarithm function of order $-1/2$. The asymptotic behavior of $\Lambda_{2,\xi}$ for $\xi\gg 1$ is,
\begin{equation}
\label{equation: lambda 2 large xi}
    \Lambda_{2,\xi} \simeq \frac{\sqrt{2\Delta}}{\pi\xi^2} e^{z^2/2}e^{-\xi^2/4} ,
\end{equation}
while for $\xi \ll 1$ it is,
\begin{equation}
\label{equation: lambda 2 small xi}
    \Lambda_{2,\xi} \simeq \sqrt{\frac{2\Delta}{\pi}}\frac{1}{\xi^3} e^{z^2/2} .
\end{equation}
Equation~(\ref{equation: lambda xi}) gives for $\Lambda_*$ in the $\mu \to 1$ limit
\begin{equation}
\label{equation: cross large xi}
    \Lambda_* \simeq 1-\frac{e^{-z^2/2}}{z\sqrt{2\pi}} \left[ 1 - \frac{1}{\sqrt{\pi}}\int_{\xi}^{\infty} \dd x\,\text{Li}_{-1/2}\left(e^{-x^2/4}\right) \right] ,
\end{equation}
so we see that the correction term is a function only of the crossover variable $\xi$. Figure~\ref{figure: lambda xi} depicts $\exp(-z^2/2)\Delta^{-1/2}\Lambda_2$, where one can see the predicted data collapse for three values of $\Delta$ (see the inset for the small- and large-$\xi$ asymptotics).
\begin{figure}
	\includegraphics[width=0.4\textwidth]{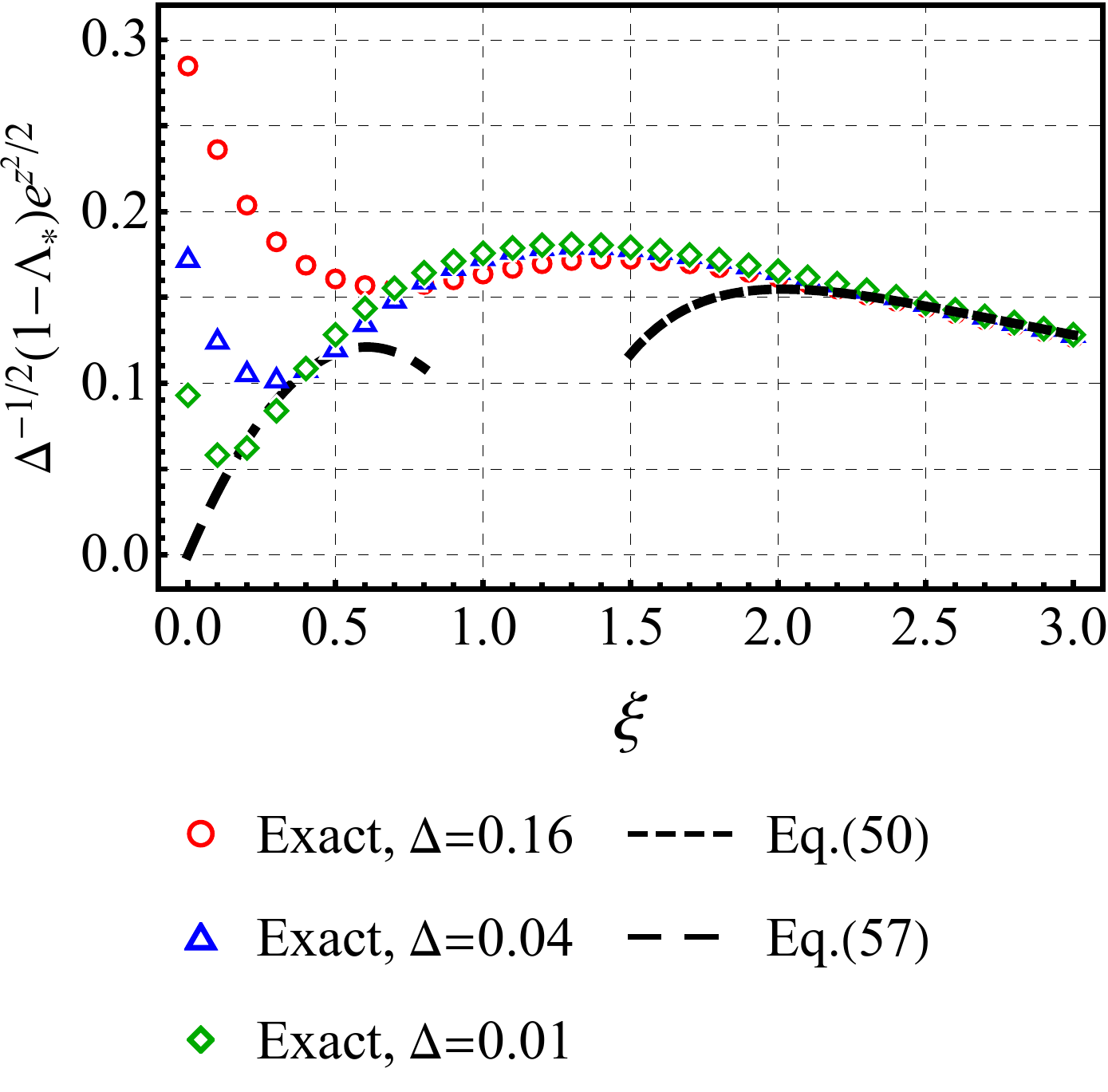}
	\caption{(Color online) {\bf The crossover region:} The scaled eigenvalue $\Delta^{-1/2}(1-\Lambda_*)\exp(z^2/2)$, obtained from numerical evacuations of Eq.~(\ref{equation: eigensystem}), is plotted as a function of the crossover variable $\xi\equiv z\sqrt{\Delta}$ for three values of $\Delta$: $0.16$, $0.04$, and $0.01$. A data collapse is seen as $\Delta\to0$. The large- and small-$\xi$ asymptotics, Eqs.~(\ref{equation: cross large xi}) and (\ref{equation: cross small xi}) respectively, are also indicated.}
\label{figure: crossover}
\end{figure}

\subsection{The $\mu \simeq 1$ regime}
\label{subsection: mu to 1}

It is also interesting to further investigate the $1-\mu \ll 1$ regime. Expanding Eq.~(\ref{equation: mu to 1 evidence}) to higher orders in $1-\mu$ indicates that here $P_*(x;z)$ obeys the continuous-sampling limit differential equation to all orders, giving rise to the parabolic cylinder function. There is, however, a boundary layer of width $\sim (1-\mu)^{1/2}$ near $z$, due to the upper limit on the integral. Therefore, for $\Delta \to 0$ and $x \to z$, let us define the inner scaled variable $s$ and function $Q(s)$ as
\begin{equation}
	s \equiv \frac{z-x}{\sqrt{2\Delta}} , \quad Q(s) \equiv \frac{P_*(x;z)}{\sqrt{\Delta}} .
\end{equation}
Importantly, for $\Delta\to 0$ one has $\Lambda_*\simeq 1-\Delta\lambda_*\to 1$, as already mentioned. Thus, to leading order Eq.~(\ref{equation: eigensystem}) becomes
\begin{equation}
\label{equation: eigensystem inner}
	Q(s) = \int_0^{\infty} \dd s' \, Q(s') \phi(s-s') .
\end{equation}
For $s\to\infty$, the lower boundary of the integral can be extended to $-\infty$, and we find that $s+\alpha$ for a constant $\alpha\in\mathbb{R}$ is the large-$s$ solution of Eq.~(\ref{equation: eigensystem inner}). Thus, the eigenfunction $Q(s)$ obeys
\begin{equation}
\label{equation: alpha definition}
	\lim_{s\to\infty} [Q(s)-s] = \alpha ,
\end{equation}
where we implicitly determined its normalization. In appendix~\ref{appendix: inner} we find the value of $\alpha$ to be
\begin{equation}
	\alpha = -\frac{\zeta(1/2)}{\sqrt{2\pi}} \approx 0.5826 ,
\end{equation}
where $\zeta(\cdot)$ is the Riemann zeta function. For the outer region, we write
\begin{equation}
\label{equation: outer eigenfunction}
	P_*(x;z) \simeq C\exp\left(-\frac{x^2}{4}\right)\text{D}_{\lambda_*+\sqrt{\Delta}\lambda_{1/2}}(-x) .
\end{equation}
Plugging $x$ in terms of $s$ into the above and expanding for small $\Delta$, we can stitch the outer region to the inner one, for which $s\to\infty$ and thus $P_*(x;z)\simeq\sqrt{\Delta}(s+\alpha)$. We find
\begin{equation}
	C = - \frac{e^{z^2/4}}{\sqrt{2}\text{D}_{1+\lambda_*}(-z)} , \quad \lambda_{1/2}(z) = \frac{\zeta(1/2)}{\sqrt{\pi}} \frac{\text{D}_{1+\lambda_*}(-z)}{\text{D}^{(0,1)}_{\lambda_*}(-z)} \mathop{\sim}_{z\to\infty} \frac{\zeta(1/2)}{\sqrt{2\pi^2}} z^2 e^{-z^2/2} ,
\end{equation}
where $\text{D}^{(0,1)}_{\lambda}(\cdot)\equiv(\dd/\dd\lambda)\text{D}_{\lambda}(\cdot)$. The results of this perturbative solution agree well with the direct numerical calculation, as seen in Figs.~\ref{figure: lambda and p for fixed z} and \ref{figure: lambda and p for large mu} for fixed $z$s and $\Delta$s, respectively. Notice also that
\begin{equation}
\label{equation: cross small xi}
    \frac{1-\Lambda_*}{\Delta} \simeq \lambda_* + \sqrt{\Delta}\lambda_{1/2} \simeq \frac{z e^{-z^2/2}}{\sqrt{2\pi}}\left[ 1 + \frac{\zeta(1/2)}{\sqrt{\pi}}\sqrt{\Delta}z\right] ,
\end{equation}
thus the correction term emerging for finite $\Delta$s is only a function of the crossover variable $\xi=\sqrt{\Delta}z$. Figure~\ref{figure: crossover} shows the scaled eigenvalue $\Delta^{-1/2}(1-\Lambda_*)\exp(z^2/2)$ as a function of the crossover variable, where one can see a data collapse in the relevant regime.

\section{A qualitative argument}
\label{section: qualitative}

The question remains, why does the IID limit dominate for large $z$ in the OU process? To understand the cause of this phenomenon, let us return to the stochastic map, Eq.~(\ref{equation: stochastic map}). Expanding for small $\Delta$, we obtain the Euler-Maruyama update equation
\begin{equation}
\label{equation: stochastic map expanded}
	x_{n+1} - x_n \simeq - \Delta x_n + \sqrt{2\Delta}\eta_n .
\end{equation}
We see from this that the displacement in $x$ is affected by a direct competition between two terms. Suppose that one has reached a given EV $z$. The probability that this EV will be crossed in the next update is
\begin{equation}
\label{equation: qualitative}
    \text{Prob}\left(x_{n+1}-x_n>0\right) \simeq \text{Prob}\left(\eta_n>\sqrt{\Delta/2}z\right) = 1 - \Phi\left(\sqrt{\frac{\Delta}{2}}z\right) ,
\end{equation}
since as mentioned, $\eta_n$ is distributed as a standard Gaussian. Thus, for any finite $\Delta$, we see that a large enough $z$ gives rise to a vanishing probability for the crossing event. Namely, the term $-\Delta x_n$ always wins, returning one back to the equilibrium location of $x=0$, which is what to be expected from an IID behavior. However, as $\Delta\to0$, the EV $z$ is crossed with probability $1/2$, independently of $z$. This means that one effectively starts at $z$ with an equal chance of going right or left, i.e. the process has a strong memory of its previous value, an indicator of the extremely correlated behavior. For small $\Delta$, the deterministic drift term becomes significant when $\Delta z\sim {\cal O}(\sqrt{\Delta})$, or $z\sqrt{\Delta}\sim {\cal O}(1)$, precisely the crossover regime we identified above.

To further explore the nature of this transition, consider the mean EV $\bar{z}$ of the OU model for a fixed but small value of $\Delta$. As we have pointed out above, since $\bar{z}$ increases with $T$, $\bar{z}$ will demonstrate a transition from being described by the continuous-sampling limit to the IID behavior as $T$ is increased. We can identify the rough magnitude of $T$ at which this transition occurs, $T_{\rm cross}(\Delta)$. Since in the limit of $T\to\infty$ the EV distribution of the continuous-sampling case converges to a Gumbel, and similarly for the IID case when $N\to\infty$, one can approximate $\bar{z} \simeq \sqrt{2\ln(T)}$ using Eq.~(\ref{equation: bn def}). Since at the transition $\bar{z} \sim {\cal O}(\Delta^{-1/2})$, we obtain that $T_{\rm cross}(\Delta)\sim\exp(1/2\Delta)$.

The argument that follows from Eq.~(\ref{equation: qualitative}) is very general, and is driven by the unbounded growth of the deterministic force term with $x$. Thus, we should expect that for all forces which approach infinity as $x\to\infty$, the IID behavior will dominate for large $z$ (and therefore for large $T$). This implies, however, that very different behavior would be expected for forces which decrease to zero as $x\to\infty$. In these cases, the noise term would dominate over the force term as $z$ becomes larger, i.e., for large $T$. Then, the EV statistics would be expected to diverge from the IID limit for large $T$ and converge instead to the continuous-sampling limit. We test this prediction in the next section, after deriving the large $T$ behavior for general forces.
\begin{figure}
	\includegraphics[width=0.75\textwidth]{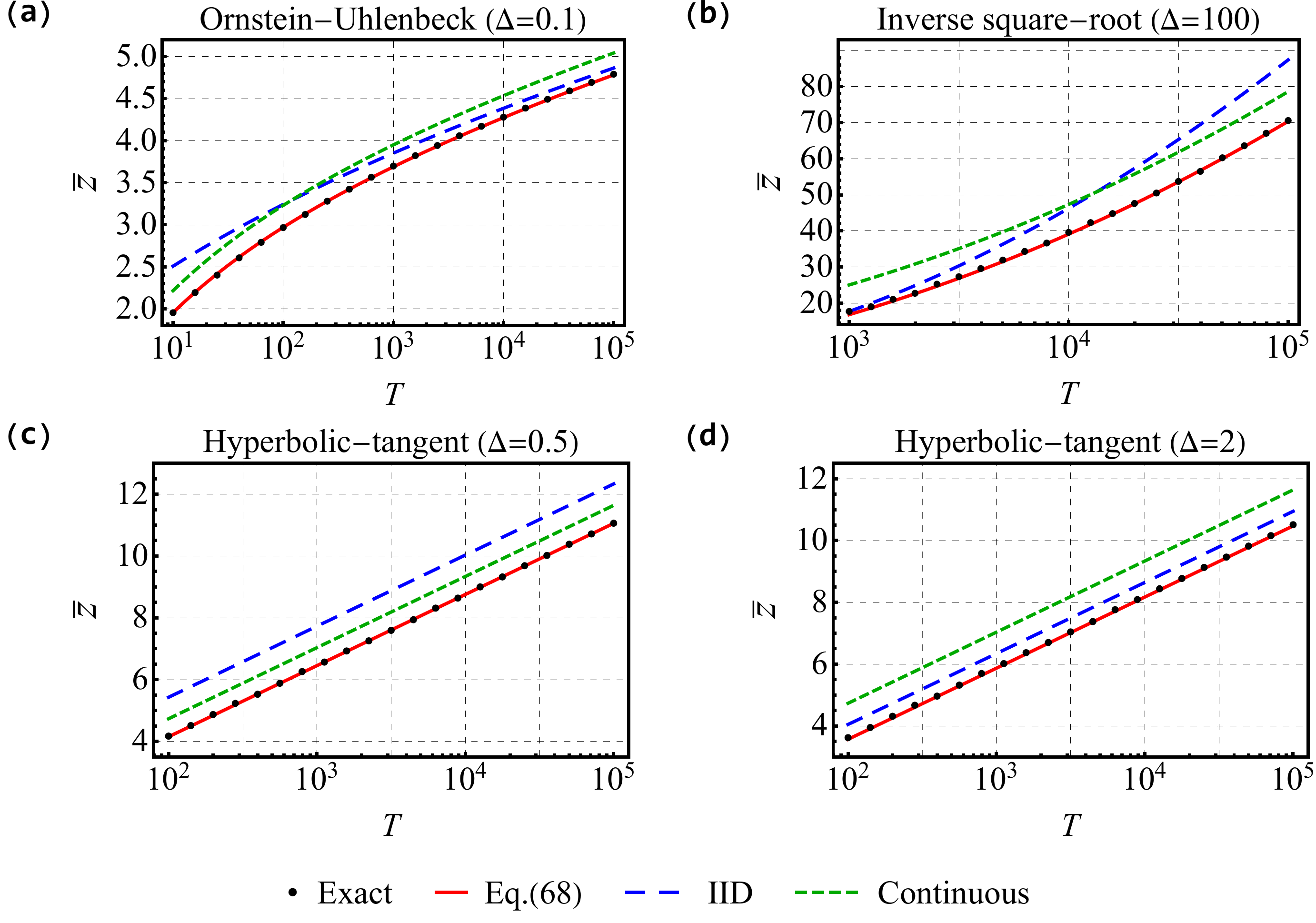}
	\caption{(Color online) {\bf The mean EV $\bar{z}$ of a correlated stochastic process $x(t)$}, evolving via variants of the Langevin equation, Eq.~(\ref{equation: langevin}), with an even spatial potential field corresponding to (a) the OU model, (b) a potential growing asymptotically as $|x|^{1/2}$, and (c,d) a hyperbolic-tangent potential. Seen are numerical evaluations for the four cases, where the sampling interval is (a) $\Delta=0.1$, (b) $\Delta=100$, (c) $\Delta=0.5$, and (d) $\Delta=2$. Also depicted are the IID case, $[F_{\rm g}^{\rm I}(z)]^{T/\Delta}$ [given by Eq.~(\ref{equation: general iid limit})], and the continuous-sampling limit, $[F_{\rm g}^{\rm c}(z)]^T$ [given right below Eq.~(\ref{equation: general iid limit})], for each scenario. For (a), the true values starts close to the continuous-sampling limit, and then crosses over to the IID description. The opposite happens for (b), as for this case the force vanishes for large distances. The borderline case is seen in (c,d), where the IID and continuous-sampling curves do not seem to intersect. Note that going to the first correction to the continuous-sampling limit, given by Eq.~(\ref{equation: bar z approx}), provides excellent predictions for $\bar{z}$. The ``exact'' means in each panel were produced by sampling $10^4$ trajectories, with an initial condition of $x=0$ for all runs. The underlying time increment for panels (b-d) was taken to be $0.001$.}
\label{figure: other cases}
\end{figure}

\section{General forces: continuous sampling}
\label{section: general}

In this section, we consider Eq.~(\ref{equation: fokker planck}) with a general potential $U(x)$,
\begin{equation}
\label{equation: fokker planck general}
    \frac{\partial P}{\partial t} = \frac{\partial^2 P}{\partial x^2} + \frac{\partial}{\partial x} \left( P \frac{\dd U}{\dd x} \right) ,
\end{equation}
corresponding to a general force of $-U'(x)$ in the Langevin equation Eq.~(\ref{equation: langevin}). We use the approach of Majumdar, et al. \cite{Majumdar}, which is in fact extremely general and can be used to derive the EV statistics for any Langevin equation with a binding potential which grows as a positive power of $x$. We have to consider separately two cases. The first is when the potential grows faster than linearly in $x$. In appendix~\ref{appendix: green 1}, we solve Eq.~(\ref{equation: fokker planck general}) in perturbation theory for a general even potential which is assumed to behave asymptotically as $U(x\to\pm\infty)\propto |x|^{\beta}$, with $\beta>1$. In this case, we find the smallest, ground state eigenvalue, to be
\begin{equation}
\label{equation: lambda eigenvalue for large z inc}
	\lambda_{\rm g}^{\beta>1}(z) \simeq \frac{1}{Z} \left[ \int_0^z\dd x\,e^{U(x)} \right]^{-1} \mathop{\sim}_{z\to\infty} \frac{1}{Z} U'(z) e^{-U(z)} ,
\end{equation}
where
\begin{equation}
\label{equation: partition function}
	Z \equiv \int_{-\infty}^{\infty} \dd x \, e^{-U(x)}
\end{equation}
is the partition function. This of course reduces to the Majumdar, et al. OU result for $U(z)=z^2/2$.

Things are more complicated when the potential grows slower than linearly, i.e. $U(x)\propto |x|^{\beta}$ with $0<\beta<1$ as $|x|\to \infty$, so that the force decays to zero for large $x$. Here, the spectrum of the Fokker-Planck equation on the semi-infinite domain $-\infty \le x \le z$ is not discrete, and the eigenvalues go continuously to $0$. Treating this case requires a very different approach, which is beyond the scope of this paper. However, if we use a reflective boundary condition at $x=0$, solving the problem of $x\in[0,z]$ instead, the spectrum is indeed discrete and we can proceed as before. We find in appendix~\ref{appendix: green 2} the smallest eigenvalue in this case to simply be $\lambda_{\rm g}^{ \beta<1}(z) = 2\lambda_{\rm g}^{\beta>1}(z)$.

Note that the effective IID underlying CDF,
\begin{equation}
\label{equation: general iid limit}
    F_{\rm g}^{\rm I}(z) = 1 - \frac{\gamma}{Z} \int_z^{\infty} \dd x\, e^{-U(x)} \mathop{\sim}_{z\to\infty} 1-\frac{\gamma}{Z}\frac{e^{-U(z)}}{U'(z)} , \quad \gamma \equiv \left\{ \begin{aligned} &1 \quad 1<\beta \\ &2 \quad 0<\beta<1 \end{aligned} \right. ,
\end{equation}
differs from its continuous-sampling effective CDF, $F_{\rm g}^{\rm c}(z) = \exp[-\gamma\lambda_{\rm g}^{\beta>1}(z)] \simeq 1-\gamma\lambda_{\rm g}^{\beta>1}(z)$, by a prefactor of $[U'(z)]^2$, proportional to $z^{2\beta-2}$ as $z\to\infty$. Thus, for $0<\beta<1$, the latter PDF decays faster than the former. As the results of discrete samplings must always lie below the continuous EV PDF curve for large $T$, one must infer that for forces which vanish with the distance, the asymptotic behavior at large-$z$ is dictated by the continuous sampling limit, in contradistinction to what happens for diverging forces, e.g. the OU model. For the case of $\beta=1$, namely an asymptotically linear potential, the $z^{2\beta-2}$ prefactor is absent. In this case, both the effective IID EV distribution and its continuous-sampling limit counterpart are asymptotically purely exponential.

This calculation for a general potential can be extended to order $\sqrt{\Delta}$, along the lines of what we did for the OU process. Indeed, by rewriting Eq.~(\ref{equation: eigensystem}) with a general potential term, one can see that the behavior inside the boundary layer discussed in subsection~\ref{subsection: mu to 1} is not affected by a change of potential to this order, and thus Eq.~(\ref{equation: eigensystem inner}) still holds. Hence, we generalize the solution of the outer region Eq.~(\ref{equation: outer eigenfunction}) to $P_{\rm g}(x;\lambda_{\rm g}+\sqrt{\Delta}\lambda^{\rm g}_{1/2})$. The eigenfunction $P_{\rm g}[x;\lambda_{\rm g}(z)]$ is associated with the eigenvalue $\lambda_{\rm g}$, and is the long-time limit solution of Eq.~(\ref{equation: fokker planck general}), obeying
\begin{equation}
	\frac{\dd^2}{\dd x^2} P_{\rm g} + \frac{\dd}{\dd x} \left[\frac{\dd U}{\dd x}P_{\rm g}\right] = -\lambda_{\rm g}(z) P_{\rm g} , \quad P_{\rm g}[z;\lambda_{\rm g}(z)] = 0 .
\end{equation}
Stitching these general inner and outer solutions, we find
\begin{equation}
    \lambda_{1/2}^{\rm g}(z) = \frac{\zeta(1/2)}{\sqrt{\pi}}\frac{P_{\rm g}^{(1,0)}\left(z;\lambda_{\rm g}\right)}{P_{\rm g}^{(0,1)}\left(z;\lambda_{\rm g}\right)} = -\frac{\zeta(1/2)}{\sqrt{\pi}} \frac{\dd \lambda_g}{\dd z} ,
\end{equation}
where the superscripts $(1,0)$ and $(0,1)$ denote partial derivatives with respect to the first and second argument of $P_{\rm g}$, respectively, and the last equation is due to the triple product rule (also known as Euler's chain rule). Thus, we see that the eigenfunction $P_{\rm g}$ does not need to be known to find the correction $\lambda_{1/2}^{\rm g}$.

To test this correction, let us first obtain a formula for the mean EV $\bar{z}$ up to order $\sqrt{\Delta}$. We denote as $f_N^{\rm g}(z)$ and $F_N^{\rm g}(z)$ the discretely-sampled general-potential PDF and CDF, respectively, and use integration by parts to get the following equation for $\bar{z}$,
\begin{equation}
    \bar{z} \equiv \int_{-\infty}^{\infty}\dd z \, f_N^{\rm g}(z) z = \int_0^{\infty} \dd z \, [1-F_N^{\rm g}(z)] - \int_{-\infty}^0 \dd z \, F_N^{\rm g}(z) .
\end{equation}
Denoting $\Lambda_{\rm g}(z)$ as the general-potential equivalent of the OU process' $\Lambda_*(z)$, the discretely-sampled CDF $F_N^{\rm g}(z)$, with $N=T/\Delta$ and $\Lambda_{\rm g} \simeq 1-\Delta(\lambda_{\rm g}+\sqrt{\Delta}\lambda_{1/2}^{\rm g})$, can be approximated for small $\Delta$ as
\begin{equation}
    F_N^{\rm d}(z) \simeq \exp\left\{-N\ln\left[\frac{1}{\Lambda_{\rm g}(z)}\right]\right\} \simeq e^{-T\lambda_{\rm g}(z)} \left[1-T\sqrt{\Delta}\lambda_{1/2}^{\rm g}(z)\right] .
\end{equation}
This gives for the mean EV
\begin{equation}
\label{equation: bar z approx}
    \bar{z} \simeq \int_0^{\infty} \dd z \left[1-e^{-T\lambda_{\rm g}(z)}\right] + \sqrt{\frac{\Delta}{\pi}} \zeta\left(\frac{1}{2}\right) ,
\end{equation}
namely the correction term does not depend on $T$. In the above calculation, we used $\lambda_{\rm g}(\infty)=0$ and $\lambda_{\rm g}(-\infty)=\infty$. Note that $\int_{-\infty}^0\dd z\,\exp[-T\lambda_{\rm g}(z)]$ is exponentially small with $T$, and so it is omitted.

Next, we visualize these results by comparing between three binding forces, differing according to their behavior at $x\to\infty$. Figure~\ref{figure: other cases} shows the mean EV $\bar{z}$ as a function of the total duration $T$, for (a) an increasing force corresponding to the OU process discussed above, (b) a force that decreases asymptotically as an inverse square-root of $x$, whose potential is $U(x)=(1+x^2)^{1/4}$, and (c,d) an asymptotically constant force derived from the potential $U(x)=\ln[\cosh(x)]$. Indeed, we see in panel (a) that $\bar{z}$ starts close to the continuous-sampling curve, and then approaches the IID curve as $T$ increases. However, for the inverse square-root potential the opposite occurs, because as heuristically explained above, when the force diminishes with the distance the continuous-sampling limit dominates the large-$z$ behavior, see panel (b). The asymptotically constant force is a borderline case, which corresponds to an exponential distributions of both the IID description and the continuous-sampling limit (up to exponentially small corrections in $z$). This is seen in panels (c) and (d), where both continuous-sampling and IID curves are parallel, and do not seem to intersect one another. The order $\sqrt{\Delta}$ results given by Eq.~(\ref{equation: bar z approx}) excellently match the simulated data. Finally, we note that the crossover point between the continuously-sampled and IID limits for the OU model as calculated above leads to $T_{\rm cross}(1/10)\approx 150$, which is reasonably close to the intersection point between
the two limiting curves of $\approx 100$.

\section{Summary}
\label{section: summary}

In this paper, we have explored the extreme statistics of correlated random variables, by analyzing the discretely sampled OU process. We showed analytically and numerically that for any non-zero sampling interval $\Delta$, the EV PDF at large times $T$ approaches the EV of $T/\Delta$ IID samples drawn from the Gaussian equilibrium distribution. We provided a simple explanation for this phenomenon based of a competition of the force and the noise terms in the OU process' Langevin equation. Exploiting this insight, we predicted that forces which decay with the distance should present an opposite behavior, namely the EV PDF should converge to that of the continuously sampled limit. We verified this prediction with an example of a force decaying as an inverse square-root of the distance. This qualitative difference between super-linear and sublinear potentials will clearly be reflected in other EV properties, such as record statistics, and are a worthy subject of future inquiry.

\begin{acknowledgments}

The support of the Israel Science Foundation via Grant No. 1898/17 is acknowledged.

\end{acknowledgments}

\appendix
\renewcommand\thefigure{\thesection\arabic{figure}}
\setcounter{figure}{0}

\section{The numerical scheme}
\label{appendix: numerics}

Here we describe the method used to numerically solve Eq.~(\ref{equation: eigensystem}). First, we split the integral as
\begin{equation}
\label{equation: numerics 1}
	\Lambda_*(z)P_*(x;z) = \int_{-\infty}^{x_{\rm m}} \dd x' \, \frac{P_*(x';z)}{\sqrt{2\pi(1-\mu^2)}} \exp\left[-\frac{(x-\mu x')^2}{2(1-\mu^2)}\right] + \int_{x_{\rm m}}^z \dd x' \, \frac{P_*(x';z) }{\sqrt{2\pi(1-\mu^2)}} \exp\left[-\frac{(x-\mu x')^2}{2(1-\mu^2)}\right] ,
\end{equation}
for some negative $x_{\rm m}$ with $|x_{\rm m}| \gg 1$, where we have removed the Heaviside step function. Note that when $z\to\infty$, Eq.~(\ref{equation: eigensystem}) can be solved in terms of the standard Gaussian $\phi(x)$ with $\Lambda_*(z)=1$, and also that taking $x\to-\infty$ has a similar mathematical consequence as taking $z\to\infty$. Thus, our next step is to express the eigenfunction as
\begin{equation}
\label{equation: numerics 2}
    P_*(x;z) \approx \left\{
    \begin{aligned}
        & P^*_n(x;z) & x_{\rm m}\le x\le z \\
        & \phi(x) & -\infty<x<x_{\rm m}
    \end{aligned}
    \right. ,
\end{equation}
where $P^*_n(x;z)$ is the $[x_{\rm m},z]$-part of the eigenfunction corresponding to the $n$th iteration. Similarly, we denote $\Lambda^*_n(z)$ as the $n$th iteration's eigenvalue. Substituting Eq.~(\ref{equation: numerics 2}) into Eq.~(\ref{equation: numerics 1}), the latter's first term can be computed explicitly, and we obtain
\begin{equation}
\label{equation: numerics 3}
	\Lambda^*_n(z)P^*_n(x;z) = \phi(x)\left\{ 1 - \frac{1}{2}\erfc\left[ \frac{x_{\rm m} - \mu x}{\sqrt{2(1-\mu^2)}} \right] \right\} + \int_{x_{\rm m}}^z \dd x' \, \frac{P^*_{n-1}(x';z)}{\sqrt{2\pi(1-\mu^2)}} \exp\left[-\frac{(x-\mu x')^2}{2(1-\mu^2)}\right] ,
\end{equation}
for $x\in[x_{\rm m},z]$. Then, assuming $P^*_{n-1}(x;z)$ is known, we discretize $x'$ on the interval $[x_{\rm m},z]$ and perform the integral of Eq.~(\ref{equation: numerics 3}). We find $\Lambda^*_n(z)$ by evaluating Eq.~(\ref{equation: numerics 3}) at $x=x_{\rm m}$, where due to continuity $P^*_n(x_{\rm m};z)=\phi(x_{\rm m})$, yielding
\begin{equation}
	\Lambda^*_n(z) = 1 - \frac{1}{2}\erfc\left( \frac{x_{\rm m}}{\sqrt{2}} \sqrt{\frac{1-\mu}{1+\mu}} \right) + \frac{1}{\phi(x_{\rm m})} \int_{x_{\rm m}}^z \dd x' \, \frac{P^*_{n-1}(x';z)}{\sqrt{2\pi(1-\mu^2)}} \exp\left[-\frac{(x_{\rm m}-\mu x')^2}{2(1-\mu^2)}\right] .
\end{equation}
Using this value, we obtain $P^*_n(x;z)$ for $x\in[x_{\rm m},z]$. Starting with $P^*_0(x;z)=\phi(x)$ and continuing to iterate gives a series of approximations to $P_*(x;z)$ which converges efficiently. We define a measure of convergence to determine the stopping point of this iterative process,
\begin{equation}
\label{equation: numerics 4}
	\mathcal{E} \equiv \left|\frac{\Lambda^*_{100m}(z)}{\Lambda^*_{100(m-1)}(z)}-1\right| , \quad 1<m\in\mathbb{N} .
\end{equation}
This prescription was used to obtain the numerical data presented in Figs.~\ref{figure: lambda and p for fixed z} and \ref{figure: lambda and p for large mu}. We used $x_{\rm m}=-5$, $\mathcal{E}=10^{-11}$, and the discretization step in $x$ was $0.002$.

While this method works well for not-so-large $z$s, working with machine-precision $\Lambda^*_n$ when $z$ is large is insufficient, since the relative change with each iteration falls below that. The solution to this problem is working with different representations of the eigenfunction and eigenvalue, based on the large-$z$ asymtotics (see subsection \ref{subsection: large z}),
\begin{equation}
    \tilde{P}_*(x;z) \equiv 1-\frac{P_*(x;z)}{\phi(x)} , \quad \tilde{\Lambda}_*(z) \equiv 1-\Lambda_*(z) .
\end{equation}
This changes Eqs.~(\ref{equation: numerics 2}-\ref{equation: numerics 4}) into
\begin{equation}
    \tilde{P}_*(x;z) \approx \left\{
    \begin{aligned}
        & \tilde{P}^*_n(x;z) & x_{\rm m}\le x\le z \\
        & 0 & -\infty<x<x_{\rm m}
    \end{aligned}
    \right. ,
\end{equation}
\begin{equation}
    \tilde{\Lambda}^*_n(z)+\left[1-\tilde{\Lambda}^*_n(z)\right]\tilde{P}^*_n(x;z) = \frac{1}{2} \erfc\left[ \frac{z - \mu x}{\sqrt{2(1-\mu^2)}} \right] + \int_{x_{\rm m}}^z \dd x' \, \frac{\tilde{P}^*_{n-1}(x';z)}{\sqrt{2\pi(1-\mu^2)}} \exp\left[-\frac{(x'-\mu x)^2}{2(1-\mu^2)} \right]
\end{equation}
for $x\in[x_{\rm m},z]$,
\begin{equation}
    \tilde{\Lambda}^*_n(z) = \frac{1}{2} \erfc\left[ \frac{z - \mu x_{\rm m}}{\sqrt{2(1-\mu^2)}} \right] + \int_{x_{\rm m}}^z \dd x' \, \frac{\tilde{P}^*_{n-1}(x';z)}{\sqrt{2\pi(1-\mu^2)}} \exp\left[-\frac{(x'-\mu x_{\rm m})^2}{2(1-\mu^2)} \right] ,
\end{equation}
and
\begin{equation}
	\tilde{\mathcal{E}} \equiv \left|\frac{\tilde{\Lambda}^*_{100m}(z)}{\tilde{\Lambda}^*_{100(m-1)}(z)}-1\right| , \quad 1<m\in\mathbb{N} .
\end{equation}
Starting with $\tilde{P}^*_0(x;z)=0$, we used this revised method to obtained the numerical data for Figs~\ref{figure: ou samples}, \ref{figure: lambda and p for small mu}, and \ref{figure: crossover}. Here, we set $x_{\rm m}=-5$ and $\tilde{\mathcal{E}}=10^{-7}$. The discretization step in $x$ was $0.01$.

\section{Finding $\alpha$ of the inner solution}
\label{appendix: inner}

In this appendix we find the parameter $\alpha$ of the inner problem of the $\mu \simeq 1$ regime, defined in Eq.~(\ref{equation: alpha definition}). Throughout this appendix, the notation of $a,b,\delta,\lambda_k$ is used to denote different objects than their main-text counterparts.

\subsection{A discretized representation}

\begin{figure}
	\includegraphics[width=0.4\textwidth]{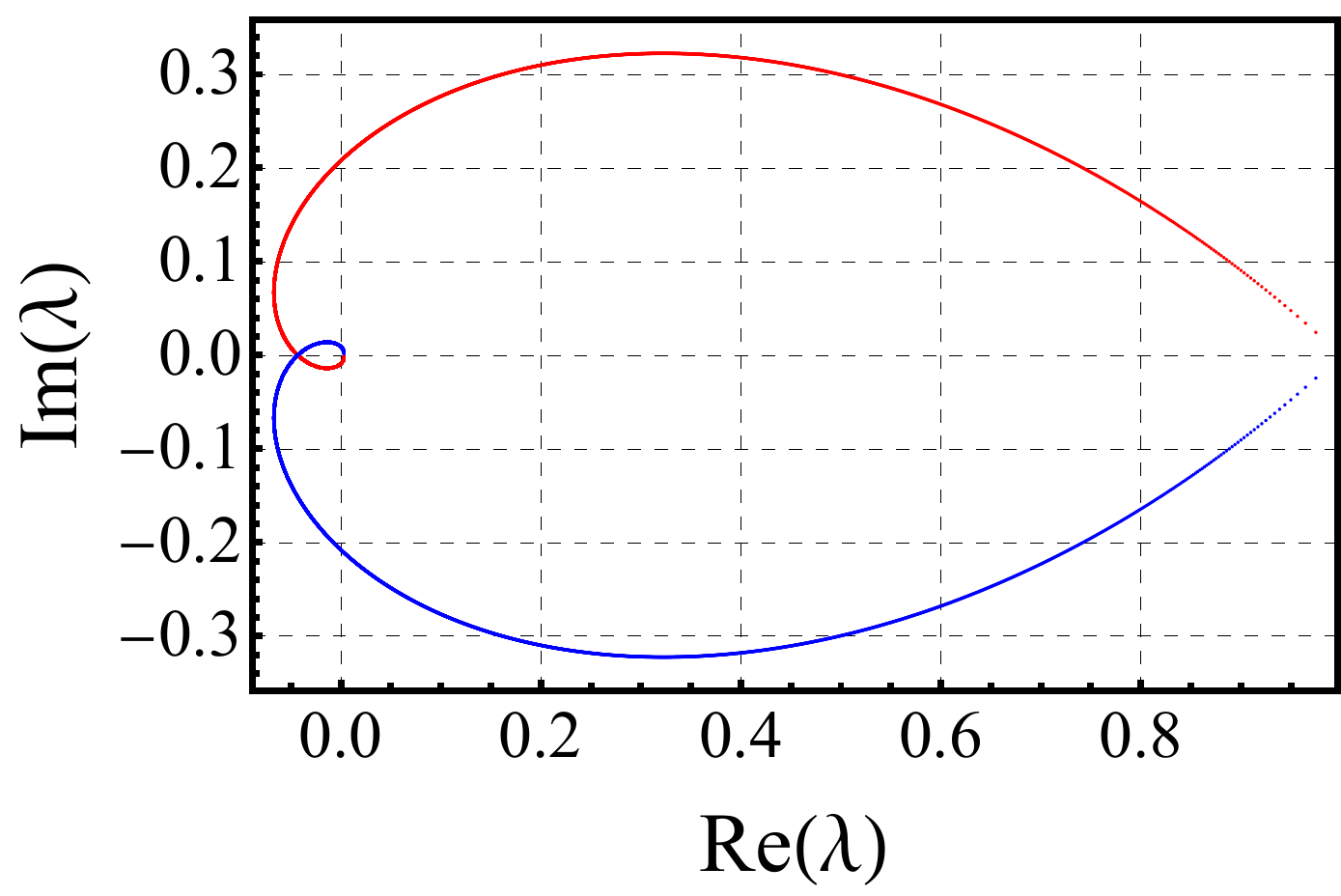}
	\caption{(Color online) {\bf The solutions of Eq.~(\ref{equation: discretizing 2}) with magnitude smaller than unity} come in conjugate pairs. Seen are the first $5\cdot 10^4$ roots given by Eq.~(\ref{equation: discretizing 3}) for $\delta=10^{-2}$, where red corresponds to $b>0$ and blue to $b<0$. The smaller $\delta$ becomes, the closer to $1$ are the two $|\lambda_1^{\pm}|$. When $\delta \to 0$, the curves become continuous.}
\label{figure: roots}
\end{figure}

We start by rewriting Eq.~(\ref{equation: eigensystem inner}) as
\begin{equation}
	Q(s) = \int_{-s}^{\infty} \dd s' \, Q(s+s') \phi(s') ,
\end{equation}
and discretizing the integral over $s'\in[-s,\infty)$ for some $\delta \ll 1$. Defining $m \equiv s/\delta$, $Q_m \equiv Q(\delta m)$, and $\phi_m \equiv \phi(\delta m)$, we get
\begin{equation}
	Q_m = \delta \sum_{m'=-\infty}^{\infty} Q_{m+m'} \phi_{m'} ,
\end{equation}
together with the boundary conditions $0=Q_{-1}=Q_{-2}=\cdots$. The general solution of this is found by substituting $Q_m \propto \lambda^m$, where $\lambda$ is an eigenvalue to be determined, which gives
\begin{equation}
\label{equation: discretizing 1}
	1 = \delta \sum_{k=-\infty}^{\infty} \phi_k \lambda^k = \exp\left[\frac{\ln^2(\lambda)}{2\delta^2}\right] \theta_3\left[-\frac{\pi\ln(\lambda)}{\delta^2} , e^{-2\pi^2/\delta^2}\right] ,
\end{equation}
where $\theta_3(\cdot,\cdot)$ is the third elliptic theta function. Since $\lambda$ is generally complex, we substitute $\ln(\lambda) = a+ib$ for some $a,b\in\mathbb{R}$, and expand for $0 < \delta \ll 1$, obtaining two conditions from the real and imaginary parts of Eq.~(\ref{equation: discretizing 1}),
\begin{equation}
\label{equation: discretizing 2}
	\exp\left(\frac{a^2-b^2}{2\delta^2}\right)\cos\left(\frac{ab}{\delta^2}\right) = 1 , \quad \sin\left(\frac{ab}{\delta^2}\right) = 0 ,
\end{equation}
yielding $a = -\sqrt{2\pi\delta^2k}$ and $b = \pm\sqrt{2\pi\delta^2k}$ with $k \in \mathbb{N}$, such that
\begin{equation}
\label{equation: discretizing 3}
	\lambda_k^{\pm} = \exp\left( -\sqrt{2\pi\delta^2k} \pm i\sqrt{2\pi\delta^2k} \right) .
\end{equation}
Note that there is a second set of solutions for which $a>0$, but these diverge for $m\to\infty$, and hence are omitted. Figure~\ref{figure: roots} depicts $\lambda_k^{\pm}$ for $k \le 5 \cdot 10^4$ and $\delta=10^{-2}$. The general solution which satisfies the aforementioned boundary conditions is given by a linear combination of the $\lambda_k^{\pm}$s,
\begin{equation}
\label{equation: discretizing 4}
	Q_m = \alpha_{\delta} + \delta m + \sum_{k=1}^{\infty} \left[ \beta_k^+\left(\lambda_k^+\right)^m + \beta_k^-\left(\lambda_k^-\right)^m \right] ,
\end{equation}
where $\alpha_{\delta}$ and the $\beta_k^{\pm}$s are determined by the boundary conditions. Note that the linear term $\delta m+\alpha_{\delta}$, which is the discretized representation of $s+\alpha$, arises due to a double unity eigenvalue for which $a=b=0$. Next, we proceed to show that $\alpha_{\delta}$ and the $\beta_k^{\pm}$s are given by
\begin{equation}
\label{equation: discretizing 5}
	\alpha_{\delta} = \delta\left[ 1+\sum_{k=1}^{\infty} \left(\frac{\lambda_k^+}{\lambda_k^+-1} + \frac{\lambda_k^-}{\lambda_k^--1} \right) \right] , \quad \beta_k^{\pm} = \frac{\delta\left(\lambda_k^{\pm}\right)^2}{1-\lambda_k^{\mp}/\lambda_k^{\pm}} \frac{1-\lambda_k^{\mp}}{1-\lambda_k^{\pm}} \prod_{l=1,l\ne k}^{\infty} \frac{1-\lambda_l^{\pm}}{1-\lambda_l^{\pm}/\lambda_k^{\pm}} \frac{1-\lambda_l^{\mp}}{1-\lambda_l^{\mp}/\lambda_k^{\pm}} .
\end{equation}

\subsection{Proving Eq.~(\ref{equation: discretizing 5})}

To prove Eq.~(\ref{equation: discretizing 5}), we reformulate the problem into one with a finite number of eigenvalues, by truncating the kernel $\phi_m$, taking care to preserve conservation of probability. For simplicity we drop the $\pm$ notation, and any indices are to be understood as going over the complete set of eigenvalues. The solution and its boundary conditions read
\begin{equation}
\label{equation: induction 1}
	Q_m = \alpha_{\delta} + \delta m + \sum_{k=1}^{M-1} \beta_k\lambda_k^m , \quad Q_{-k} = 0 \text{ for any integer } 1 \le k \le M ,
\end{equation}
where $M \ge 3$ is an integer. Our claim is that Eq.~(\ref{equation: induction 1}) dictates
\begin{equation}
\label{equation: induction 2}
	\alpha_{\delta} = \delta \left(1+\sum_{k=1}^{M-1}\frac{\lambda_k}{\lambda_k-1}\right) , \quad \beta_k = \frac{\delta\lambda_k^2}{1-\lambda_k} \prod_{l=1,l\ne k}^{M-1} \frac{1-\lambda_l}{1-\lambda_l/\lambda_k} ,
\end{equation}
and we shall prove it by mathematical induction. We start by setting $M=3$, such that one gets out of the boundary conditions,
\begin{equation}
	Q_{-1} = \alpha_{\delta} - \delta + \frac{\beta_1}{\lambda_1} + \frac{\beta_2}{\lambda_2} = 0 , \quad Q_{-2} = \alpha_{\delta} - 2\delta + \frac{\beta_1}{\lambda_1^2} + \frac{\beta_2}{\lambda_2^2} = 0 , \quad Q_{-3} = \alpha_{\delta} - 3\delta + \frac{\beta_1}{\lambda_1^3} + \frac{\beta_2}{\lambda_2^3} = 0 ,
\end{equation}
which are solved to yield
\begin{equation}
	\alpha_{\delta} = \delta\left(1+\frac{\lambda_1}{\lambda_1-1}+\frac{\lambda_2}{\lambda_2-1}\right) , \quad \beta_1 = \frac{\delta\lambda_1^2}{1-\lambda_1} \frac{1-\lambda_2}{1-\lambda_2/\lambda_1} , \quad \beta_2 = \frac{\delta\lambda_2^2}{1-\lambda_2} \frac{1-\lambda_1}{1-\lambda_1/\lambda_2} .
\end{equation}
Next, we assume validity for a certain $M = M_* \ge 3$, namely that the $\alpha_{\delta}$ and $\beta_k$s which are dictated by the boundary conditions of Eq.~(\ref{equation: induction 1}) are given by Eq.~(\ref{equation: induction 2}). Finally, we prove correctness for $M = M_*+1$. The boundary conditions then give $M_*+1$ equations for $1 \le j \le M_*+1$,
\begin{equation}
	Q_{-j} = \alpha_{\delta} - j\delta + \sum_{k=1}^{M_*-1}\frac{\beta_k}{\lambda_k^j} + \frac{\beta_{M_*}}{\lambda_{M_*}^j} = 0 ,
\end{equation}
Let us create $M_*$ new equations as $Q_{-j}' \equiv Q_{-j}-\lambda_{M_*}Q_{-j-1}$, and get
\begin{equation}
	Q_{-j}' = \alpha_{\delta}(1-\lambda_{M_*}) + \delta\lambda_{M_*} - j\delta(1-\lambda_{M_*}) + \sum_{k=1}^{M_*-1}\frac{\beta_k}{\lambda_k^j}\left(1-\frac{\lambda_{M_*}}{\lambda_k}\right) = 0 ,
\end{equation}
for $1 \le j \le M_*$. Defining
\begin{equation}
	\alpha_{\delta}' = \alpha_{\delta}(1-\lambda_{M_*}) + \delta\lambda_{M_*} , \quad \delta' = \delta(1-\lambda_{M_*}) , \quad \beta_k' = \beta_k\left(1-\frac{\lambda_{M_*}}{\lambda_k}\right) ,
\end{equation}
for $1 \le k \le M_*-1$, we obtain
\begin{equation}
	Q_{-j}' = \alpha_{\delta}' - j\delta' + \sum_{k=1}^{M_*-1} \frac{\beta_k'}{\lambda_k^j} = 0 .
\end{equation}
Invoking the induction assumption, we know that these equations yield
\begin{equation}
	\alpha_{\delta}' = \delta' \left(1+\sum_{k=1}^{M_*-1}\frac{\lambda_k}{\lambda_k-1}\right) , \quad \beta_k' = \frac{\delta'\lambda_k^2}{1-\lambda_k} \prod_{l=1,l\ne k}^{M_*-1} \frac{1-\lambda_l}{1-\lambda_l/\lambda_k} ,
\end{equation}
from which is it easy to see that
\begin{equation}
\label{equation: induction 3}
	\alpha_{\delta} = \delta \left(1+\sum_{k=1}^{M_*}\frac{\lambda_k}{\lambda_k-1}\right) , \quad \beta_k = \frac{\delta\lambda_k^2}{1-\lambda_k} \prod_{l=1,l\ne k}^{M_*} \frac{1-\lambda_l}{1-\lambda_l/\lambda_k} ,
\end{equation}
for $1 \le k \le M_*-1$. We are thus left with showing that $\beta_{M_*}$ also follows Eq.~(\ref{equation: induction 3}). We do so by defining a new set of $M_*$ equations as $Q_{-j}^{(1)} \equiv Q_{-j}-Q_{-j-1}$, obtaining
\begin{equation}
	Q_{-j}^{(1)} = \delta + \frac{\beta_1}{\lambda_1^j}\left(1-\frac{1}{\lambda_1}\right) + \sum_{k=2}^{M_*}\frac{\beta_k}{\lambda_k^j}\left(1-\frac{1}{\lambda_k}\right) = 0 ,
\end{equation}
for $1 \le j \le M_*$. Defining yet another set of $M_*-1$ equations as $Q_{-j}^{(2)} \equiv Q_{-j}^{(1)}-\lambda_1 Q_{-j-1}^{(1)}$, we get
\begin{equation}
	Q_{-j}^{(2)} = \delta(1-\lambda_1) + \frac{\beta_2}{\lambda_2^j} \left(1-\frac{\lambda_1}{\lambda_2}\right) \left(1-\frac{1}{\lambda_2}\right) + \sum_{k=3}^{M_*}\frac{\beta_k}{\lambda_k^j} \left(1-\frac{\lambda_1}{\lambda_k}\right) \left(1-\frac{1}{\lambda_k}\right) = 0 ,
\end{equation}
for $1 \le j \le M_*-1$. If one continues to iterate, one has for the $M_*$th iteration, $Q_{-1}^{(M_*)} \equiv Q_{-1}^{(M_*-1)}-\lambda_{M_*-1} Q_{-2}^{(M_*-1)}$, a single equation for $j=1$,
\begin{equation}
	Q_{-1}^{(M_*)} = \delta\prod_{l=1}^{M_*-1}(1-\lambda_l) + \frac{\beta_{M_*}}{\lambda_{M_*}} \left(1-\frac{1}{\lambda_{M_*}}\right) \prod_{l=1}^{M_*-1} \left(1-\frac{\lambda_l}{\lambda_{M_*}}\right) = 0 ,
\end{equation}
which is solved to yield exactly Eq.~(\ref{equation: induction 3}) with $k=M_*$. Hence, our proof is concluded.

\subsection{Back to the continuum representation}

Finally, let us calculate the value of $\alpha_{\delta}$ in the continuum representation of $\delta\to 0$. Using Eq.~(\ref{equation: discretizing 3}), we find from Eq.~(\ref{equation: discretizing 5}) that
\begin{equation}
	\alpha_{\delta} = \delta\left[1+\sum_{k=1}^{\infty} \frac{2\exp\left(-2\sqrt{2\pi\delta^2k}\right) - 2\exp\left(-\sqrt{2\pi\delta^2k}\right) \cos\left(\sqrt{2\pi\delta^2k}\right)}{\exp\left(-2\sqrt{2\pi\delta^2k}\right) - 2\exp\left(-\sqrt{2\pi\delta^2k}\right) \cos\left(\sqrt{2\pi\delta^2k}\right)+1} \right] .
\end{equation}
Note that in the aforementioned limit, this sum cannot be naively changed into an integral due to the discreteness of the first terms. Thus, we break the sum at an arbitrary location $1/\delta^2\gg K\gg1$. For $1 \le k \le K-1$, taking the limit $\delta \to 0$ gives
\begin{equation}
	\lim_{\delta \to 0} \delta \sum_{k=1}^{K-1} \frac{2\exp\left(-2\sqrt{2\pi\delta^2k}\right) - 2\exp\left(-\sqrt{2\pi\delta^2k}\right)\cos\left(\sqrt{2\pi\delta^2k}\right)}{\exp\left(-2\sqrt{2\pi\delta^2k}\right) - 2\exp\left(-\sqrt{2\pi\delta^2k}\right)\cos\left(\sqrt{2\pi\delta^2k}\right)+1} = - \sum_{k=1}^{K-1} \frac{1}{\sqrt{2\pi k}} .
\end{equation}
For $k \ge K$, we approximate the sum with an integral, changing variables to $\kappa\equiv\sqrt{2\pi\delta^2k}$,
\begin{align}
	&\lim_{\delta\to 0} \delta \sum_{k=K}^{\infty} \frac{2\exp\left(-2\sqrt{2\pi\delta^2k}\right) - 2\exp\left(-\sqrt{2\pi\delta^2k}\right)\cos\left(\sqrt{2\pi\delta^2k}\right)}{\exp\left(-2\sqrt{2\pi\delta^2k}\right) - 2\exp\left(-\sqrt{2\pi\delta^2k}\right)\cos\left(\sqrt{2\pi\delta^2k}\right)+1} \\
	= &\lim_{\delta\to 0} \int_{\sqrt{2\pi\delta^2K}}^{\infty} \frac{\dd\kappa\,\kappa}{\pi \delta} \frac{2e^{-2\kappa}-2e^{-\kappa}\cos(\kappa)}{e^{-2\kappa}-2e^{-\kappa}\cos(\kappa)+1} = - \lim_{\delta\to 0} \int_0^{\sqrt{2\pi\delta^2K}} \frac{\dd\kappa\,\kappa}{\pi \delta} \frac{2e^{-2\kappa}-2e^{-\kappa}\cos(\kappa)}{e^{-2\kappa}-2e^{-\kappa}\cos(\kappa)+1} = \sqrt{\frac{2K}{\pi}} \nonumber ,
\end{align}
where in the second to last transition we used $\int_0^{\infty}\dd\kappa\,\kappa[e^{-2\kappa} - e^{-\kappa}\cos(\kappa)]/[e^{-2\kappa}-2e^{-\kappa}\cos(\kappa)+1] = 0$. Thus, we get
\begin{equation}
	\alpha = \lim_{\delta\to 0} \alpha_{\delta} = \lim_{K\to\infty} \left( \sqrt{\frac{2K}{\pi}} - \sum_{k=1}^{K-1} \frac{1}{\sqrt{2\pi k}} \right) = -\frac{\zeta(1/2)}{\sqrt{2\pi}} ,
\end{equation}
where $\zeta(\cdot)$ is the Riemann zeta function. Finally, presentation of the complete inner solution in Fig.~\ref{figure: lambda and p for fixed z} was done numerically. We replaced the infinite limits of summation/multiplication in Eqs.~(\ref{equation: discretizing 4}) and (\ref{equation: discretizing 5}) with some large value $M$, and evaluated $Q_m$ using a finite $\delta$. The smaller $\delta$ is, the larger $M$ must be. The aforementioned figure was plotted using $\delta=0.005$ and $M=10^7$.

\section{Solving the Fokker-Planck equation}

\subsection{For diverging forces ($\beta>1$)}
\label{appendix: green 1}

Consider a general one-dimensional stochastic process with the Fokker-Planck equation Eq.~(\ref{equation: fokker planck general}), with $U(x)$ a spatial potential field. Solving it using separation of variables yields $P(x,t) = X(x)\exp(-\lambda t)$ for some eigenvalue $\lambda>0$, where $X(x)$ obeys
\begin{equation}
\label{equation: green 1}
	\frac{\dd^2X}{\dd x^2} + \frac{\dd}{\dd x} \left( X \frac{\dd U}{\dd x} \right) + \lambda X = 0 .
\end{equation}
Let us assume that the potential $U(x)$ is an even function with an asymptotic behavior of $U(x\to\pm\infty) \propto |x|^{\beta}$, where $\beta>1$. We solve Eq.~(\ref{equation: green 1}) over the domain $x\in(-\infty,z]$, with boundary conditions $X(-\infty)=X(z)=0$ which yield $\lambda$ as a function of $z$. We do so by using perturbation theory around $\lambda=0$, for which $z\to\infty$. For the zeroth order, Eq.~(\ref{equation: green 1}) reads
\begin{equation}
	\frac{\dd^2X_0}{\dd x^2} + \frac{\dd}{\dd x} \left( X_0 \frac{\dd U}{\dd x} \right) = 0 ,
\end{equation}
whose general solution is
\begin{equation}
	X_0(x) = C_1 y_1(x) + C_2 y_2(x)
\end{equation}
where
\begin{equation}
	y_1(x) = e^{-U(x)} , \quad y_2(x) = y_1(x) \int_0^x \dd \chi \, e^{U(\chi)} ,
\end{equation}
with boundary conditions of $X_0(-\infty)=X_0(\infty)=0$. Since $y_2(x)$ decays algebraically when $x\to\pm\infty$, More precisely $y_2(x) \propto |x|^{1-\beta}$, it needs to be discarded, as the solution should approach zero for $x\to-\infty$ in an exponential manner. Therefore, we have
\begin{equation}
\label{equation: green 2}
	X_0(x) = C_1 y_1(x)
\end{equation}
as the zero-order solution. As we shall see, $|\lambda|\ll 1$, hence we use perturbation theory. Writing $X(x)=X_0(x)+X_1(x)$, we obtain the inhomogeneous equation
\begin{equation}
\label{equation: green 3}
	\frac{\dd^2X_1}{\dd x^2} + \frac{\dd}{\dd x} \left( X_1 \frac{\dd U}{\dd x} \right) = -\lambda X_0 .
\end{equation}
By the method of variation of parameters, the general solution of Eq.~(\ref{equation: green 3}) is given by
\begin{equation}
\label{equation: green 4}
	X_1(x) = \lambda C_1 \left[ y_1(x)\int_{x_1}^x\dd\chi\,y_2(\chi) - y_2(x)\int_{x_2}^x\dd\chi\,y_1(\chi) \right] ,
\end{equation}
where $x_1$ and $x_2$ are arbitrary constants. As mentioned, the decay at $x\to-\infty$ should be exponential, hence the coefficient of $y_2(x)$ must vanish in this limit. Thus, we must choose $x_2=-\infty$, and therefore
\begin{equation}
	X(x) = C_1y_1(x) + \lambda C_1 \left[ y_1(x)\int_{x_1}^x\dd\chi\,y_2(\chi) - y_2(x)\int_{-\infty}^x\dd\chi\,y_1(\chi) \right] .
\end{equation}
Setting this to zero at $x=z$ and taking $z\to\infty$ yields Eqs.~(\ref{equation: lambda eigenvalue for large z inc}) and (\ref{equation: partition function}).

\subsection{For vanishing forces ($0<\beta<1$)}
\label{appendix: green 2}

This time, we solve Eq.~(\ref{equation: green 1}) over the domain $x\in[0,z]$, where the boundary conditions are $X'(0)=X(z)=0$. The zero-order has the same general solution, and its boundary conditions read $X_0'(0)=X_0(\infty)=0$. Since $y_2'(0)=1$, we have the same solution for the zero-order, namely Eq.~(\ref{equation: green 2}). Hence, we obtain the same inhomogeneous equation for the first-order, Eq.~(\ref{equation: green 3}), solved via the method of variation of parameters to yield Eq.~(\ref{equation: green 4}). However, this time due to the boundary condition at $0$ and given that $y_2'(0)=1$, we must choose $x_2=0$. Then, we have
\begin{equation}
	X(x) = C_1y_1(x) + \lambda C_1 \left[ y_1(x)\int_{x_1}^x\dd\chi\,y_2(\chi) - y_2(x)\int_0^x\dd\chi\,y_1(\chi) \right] ,
\end{equation}
where setting this to zero at $x=z$ and taking $z\to\infty$ yields $\lambda_{\rm g}^{\beta<1}(z) = 2\lambda_{\rm g}^{\beta>1}(z)$.


\begin{thebibliography}{99}
\bibitem{Hermite2}
One can prove Eq.~(\ref{equation: hermite expansion 2}) by differentiating it with respect to $\mu$, and using $\He_n(x)=x\He_{n-1}(x)-\He_{n-1}'(x)$ together with Eq.~(\ref{equation: hermite expansion 1}) to show that the resulted expression holds. Verifying that Eq.~(\ref{equation: hermite expansion 2}) is satisfied for $\mu=0$ concludes the proof.
\end{thebibliography}
\end{document}